\documentclass[preprint2,tighten]{aastex6}

\slugcomment{The Astrophysical Journal, 24pp, 2017}
\shorttitle{Modified Disk Model}    
\shortauthors{Godon et al.}
\watermark{arXiv astroph e-print Copy}
\setwatermarkfontsize{0.6in}

\begin{document}

\title{{\bf Modifying the Standard Disk Model for the Ultraviolet 
Spectral Analysis of Disk-dominated Cataclysmic Variables. 
I. The Novalikes MV Lyrae, BZ Camelopardalis, and V592 Cassiopeiae
\altaffilmark{1}
}}

\author{Patrick Godon \altaffilmark{2}, Edward M. Sion}
\affil{Astrophysics and Planetary Science, Villanova University, 
800 Lancaster Ave., 
Villanova, PA 19085, USA}
\email{patrick.godon@villanova.edu ; edward.sion@villanova.edu }

\author{\c{S}\"olen Balman} 
\affil{Department of Physics, 
Middle East Technical University,
Ankara, Turkey}
\email{solen@astroa.physics.metu.edu.tr}   

\and

\author{William P. Blair} 
\affil{Henry A. Rowland Department of Phyiscs \& Astronomy, 
The Johns Hopkins University, Baltimore, MD 21218, USA} 
\email{wpb@pha.jhu.edu} 

\altaffiltext{1}{This e-print copy was not formatted nor edited for The
Astrophysical Journal, the figures have been reduced to span only
one column.} 

\altaffiltext{2}{Visiting in the Henry A. Rowland Department of Physics
\& Astronomy, The Johns Hopkins University, Baltimore, MD 21208}

\begin{abstract}    

The standard disk is often inadequate to model disk-dominated 
cataclysmic variables (CVs) and generates a spectrum that is bluer  
than the observed UV spectra \citep{pue07}. 
X-ray observations of these systems reveal an optically thin 
boundary layer (BL) expected to appear as an inner hole in the disk.  
Consequently, we truncate the inner disk.
However, instead of removing the inner disk,  
we impose the no-shear boundary 
condition at the truncation radius, thereby
lowering the disk temperature  
and generating a spectrum that better fits the UV data. 
With our modified disk, we analyze the archival UV spectra 
of three novalikes that cannot be fitted with standard disks. 
For the VY Scl systems MV Lyr and BZ Cam, 
we fit a hot inflated white dwarf 
WD with a cold modified disk ($\dot{M} \sim $  a few $10^{-9}M_{\odot}$/yr).   
For V592 Cas, the slightly modified disk 
($\dot{M} \sim 6 \times 10^{-9}M_{\odot}$/yr) completely dominates the UV.  
These results are consistent with  
{\it Swift} X-ray observations of these systems \citep{bal14},
revealing BLs merged with ADAF-like flows and/or hot coronae, 
where the advection of energy is likely launching an outflow  
and heating the WD, thereby explaining the high WD temperature 
in VY Scl systems.   
This is further supported by the fact that the X-ray hardness ratio 
increases with the shallowness of the UV slope in a small CV sample
we examine. Furthermore, for 105 disk-dominated systems, the 
{\it International Ultraviolet Explorer} ({\it IUE} 
spectra UV slope  decreases in the same order as the ratio of the
X-ray flux to optical/UV flux: from SU UMa's, to U Gem's, Z Cam's, UX UMa's, 
and VY Scl's. 

\end{abstract} 

\keywords{accretion, acretion disks --- novae, cataclysmic variables --- white dwarfs}

\section{{\bf Introduction: Ultraviolet Disk Spectra and 
Optically Thin Hard X-Ray in Novalikes}}   

The properties of accretion disks in binary systems 
are best studied in the brightest types of non-magnetic cataclysmic variables 
(CVs):  the novalikes (NLs) in high state 
and dwarf novae (DNe) in outburst.
These systems are compact binaries with a white dwarf (WD) 
accreting matter and angular momentum via a disk from a low-mass 
companion filling its Roche lobe. Dwarf nova systems are 
found mostly in a state of low mass accretion rate (quiescence)
and have periods of high mass accretion rate (outburst) when the disk
dominates the UV \citep{hac93}.  
NLs are found mostly in a state of high mass accretion 
in which the UV emission is predominantly
originating from the disk \citep{lad90,lad91}.
For that reason, the DNe in outburst and NLs in high state are of special
interest because they are expected to have an accretion disk
radial temperature profile given by the analytical expression 
$T_{\rm eff}(R)$ of the standard disk model \citep{pri81}.

However, 30 years ago already, \citet{wad84,wad88} showed,  
using 
{\it International Ultraviolet Explorer ({\it IUE})} 
spectra, that some CV systems 
systematically disagree with the standard
disk model when either black body or Kurucz stellar models 
were used to represent the accretion disk. A better
model was developed by \citet{hub90} explicitly including the calculation of
synthetic spectra using the standard disk model (the code TLUSTY;
\citet{wad98}).   This modeling was more successful 
and has been used since in standard procedures to assess the
mass accretion rate of systems in a state of high mass accretion rate 
(e.g.  \citet{kni02,ham07}).             
With time, however, 
it appeared that a large fraction of disk systems could  {\bf not} 
be modeled using the standard disk temperature
profile, and instead the temperature of the disk had to be modified,  
e.g. IX Vel, QU Car,  RW Sex; 
\citet{lin07,lin08,lin10}. 
A statistical study \citep{pue07} of 
33 disk systems indicated that 
a revision of the temperature profile, at least in 
the innermost part of the disk, is required: 
a large fraction of the theoretical spectra were too 
blue compared to the observed spectra.  
Some of the UV spectral modelings of disk-dominated systems 
that were in better agreement with the observations 
consisted of truncating the hot inner disk
\citep{lin05}. By truncating the the hottest
region in the disk, the slope of the UV continuum became shallower 
as the spectrum became redder. 
In her statistical analysis of {\it IUE} spectra of CVs, \citet{lad91} 
already pointed out that dwarf novae in outburst have a continuum flux level 
increasing toward the shorter wavelength (blue),  
while the continuum slope is shallower for  non-magnetic NLs and 
it is almost flat for intermediate
polars (IPs - systems in which the inner disk is expected to be 
truncated by the WD magnetic field).  
\citet{lad91} showed that {\it IUE} spectra of IPs were 
similar in slope/color to theoretical spectra of magnetically 
truncated disks with a truncation radius $\approx 5R_{\rm wd}$,  
while in contrast, a large number of DNe in outburst have been sucessfully 
modeled with non-truncated standard disk spectra \citep{ham07}.  
This suggests that the disk spectra of non-magnetic NLs,
showing a UV continuum slope intermediate between IPs and DNe
in outburst, might have a truncated disk with a truncation radius 
smaller that in IPs.

In addition to the above problem in the UV spectroscopy 
of disk-dominated systems,               
X-ray data of CVs have also been in strong disagreement with 
the theoretical expectations for more than three decades \citep{fer82}.
The culprit has been the difficult to study boundary layer (BL) 
between the accretion disk and the WD.   
About half of the disk accretion energy (in the form of kinetic energy) 
is expected to be dissipated in the BL 
between the Keplerian disk and slowly rotating stellar surface 
\citep{pri81}. 
Because of its small size, the BL 
was predicted to emit in the X-ray band:
at low accretion rates (when the WD is dominant in the
UV), the BL was expected to be optically thin and emit hard X-rays
\citep{pri79,tyl82,pop93}; 
at large accretion rates, typical of NLs in a high 
state and DNe in outburst, the BL was expected to be optically thick 
and emit soft X-rays \citep{pri77,pop93,pop95}. 
Systems in the low state indeed reveal optically thin hard X-ray emission
\citep{szk02,pan05,muk09}. 
However, systems in a state of high mass accretion often do not show 
an optically
thick soft X-ray component; instead, many exhibit optically 
thin hard X-ray emission 
\citep{pat85a,pat85b,mau95,van96,bas05,bal14}, 
with an X-ray luminosity much smaller than expected, 
i.e. much smaller than the disk luminosity. 
While optically thin hard X-ray emission from high mass accretion
rate systems was unexpected,  
it is, however, not especially inconsistent with the theoretical
work: optically thin boundary layers can occur in high mass 
accretion rate systems, since the transition to optically thin depends
not only on the mass accretion rate, but also on the WD mass, the
WD rotation rate and the (unknown) alpha viscosity parameter \citep{pop95}. 
Simulations of optically thin BLs  
\citep{pop93,pop99} show that the inner edge of the Keplerian 
(and optically thick) 
disk starts at an actual  radius $R_0 = R_{\rm wd} + \delta_{\rm BL}$, 
where the size $\delta_{\rm BL}$ of the BL is of the order of the
stellar radius $R_{\rm wd}$: $\delta_{\rm BL} \propto R_{\rm wd}$ (the BL is actually
geometrically thick). 
The direct consequence of having an optically thin and geometrically 
thick BL is that the optically thick Keplerian disk will appear to have  
an inner hole of size $\delta_{\rm BL}$ (possibly of the order of the radius
of the WD $R_{\rm wd}$). 
Two decades ago, it had already been pointed out
that optically thin BL can explain the inner hole observed in circumstellar
disks around young stellar objects (T Tauri stars, \citet{god96}).
In other words, optically 
thin boundary layer are consistent not only with the X-ray data,
but also with the UV data, as truncated optically thick disks produce a
UV continuum with a shallow slope in better agreement with the UV observations 
than non-truncated disks.

It appears  that the high mass accretion rate NL  
systems seem to systematically exhibit both optically thin hard X-ray emission 
and a UV continuum with a shallow slope that 
cannot be fitted with standard disk models,   
e.g. MV Lyr \citep{lin05,bal14}, IX Vel \citep{lon94,lin07},   
BZ Cam \citep{bal14,pri00}, V592 Cas \citep{bal14,hoa09}, 
and RW Sex \citep{cor81,lin10}.
Such a behavior indicates that 
this might be the rule for NL systems as a whole.  
We suspect that NLs, due to the optically thin BL, 
have a truncated inner disk with a radius
of truncation $\approx 2R_{\rm wd}$.  
Our recent {\it Swift} X-ray observations of the three NLs BZ Cam,
MV Lyr, and V592 Cas \citep{bal14} 
show optically thin boundary layers merged
with advection-dominated accretion flows (ADAF) and/or hot coronae. 
In such  systems, the BL energy is  
advected inwards as optically thin BLs cannot cool efficiently 
\citep{abr95}.  
The heating of the stellar surface can further result in an 
increased/inflated stellar radius.  

Motivated by this, we decided to carry out an improved UV spectral
analysis of the three NLs BZ Cam, MV Lyr, and V592 Cas, by
modifying (truncating) the standard disk model. 
The three NL systems are reviewed in the next section together 
with their archival {\it FUSE} and {\it IUE} spectra.  
The modification of the standard disk model, 
presented in section 3,
consists mainly in taking the inner radius of the optically thick disk to be 
at $R_0= R_{\rm wd} + \delta_{\rm BL}$, 
where $R_{\rm wd}$ is the non-zero temperature WD radius and 
$\delta_{\rm BL}$ is the size of the optically thin BL (which is now also
a free parameter of size $R_{\rm wd}$ or smaller). 
We caution that {\it generating a standard disk model with an inner disk radius 
$R_0 > R_{\rm wd}$ is not the same as truncating at $R_0$ a standard disk
model generated with an inner disk radius at $R_{\rm wd}$}, because 
the inner boundary condition is changed and, therefore,  generates a
new {\it mathematical} solution that has a lower temperature 
(although similar to the standard disk model, it is different).    
We briefly present the stellar spectral modeling in section 4,
and show how the disk spectra are affected by our modification. 
In section 5, the results  show that the modified disk
model provides a better fit to the observed UV spectra of these three
novalikes.  
In section 6, we check a possible correlation between the slope 
of the UV spectrum and the X-ray hardness ratio for a sample
of CVs.  
We also discuss the possibility that the optically thin 
(truncated) inner region of the  disk might be extended   
in the VY Scl subclass of NLs, and we also quantitatively check 
the departure of UV spectra from the standard disk model  
for disk-dominated CV subtypes. 
We present our conclusion in section 7. 
 
\section{{\bf The Three Novalikes MV Lyr, BZ Cam, \& V592 Cas}}  

The three systems MV Lyr, BZ Cam, and V592 Cas  
were the three targets of our {\it Swift} X-ray observations
\citep{bal14}, and while this might seem to be the main reason 
for choosing  them for our UV spectral analysis, 
these systems have some characteristics that make them especially 
relevant for the kind of spectral analysis we are performing here.   
In order to demonstrate that some NL systems 
are best modeled with truncated disks, we choose systems 
in which the shallow slope of the UV flux continuum is incontrovertible.
Namely, we want to ensure that the shallow 
slope is not due to an unknown reddening 
or to the wrong choice of system parameters such as the WD mass, radius,
the distance to the system, or its inclination. 

With a spectral coverage from $\sim$910\AA\ to $\sim$1190\AA\ ,
{\it FUSE} covers the Lyman series and is therefore more
useful in helping determine the hot component 
(WD temperature and/or mass accretion rate) of 
CVs than any other UV telescope. 
However, we wish to use    
the broadest spectral UV coverage possible, and we try to choose
systems that have both {\it FUSE} and {\it IUE} spectra, thereby 
extending from the Lyman limit $\sim 910$\AA\ 
to the near ultraviolet $\sim 3,200$\AA .  
There are many hundreds (if not thousands) of known CVs, 
and almost 300 of them have 
good {\it IUE} spectra from $\sim 1150$\AA\ 
to $\sim$3200\AA\ . 
Usually the {\it IUE} spectra are easily used to derive the reddening 
(E(B-V)) toward the system using the 2175\AA\ 
feature (with an accuracy of about 25\%, \citet{fit99}).    
Since it is not clear that one can use the same standard reddening law
for all CVs in the Galaxy (e.g. \citet{fit99}),    it would be preferable to choose 
systems with zero or negligibly small reddening $E(B-V) \approx 0$.  
In the present work we use the extinction curve from \citet{fit07}
with R=3.1 to deredden the UV spectra.  
   
Another desirable feature is the inclination
of the systems. A high inclination system will be more affected by
the orbital phase. In addition the standard disk model is a two 
dimensional flat disk and its departure from a three-dimensional
realistic disk will be more pronounced at higher inclinations. 
We therefore prefer to choose systems with a small inclination. 
   
When modeling the accretion disk, 
it is preferable to know the distance and/or WD mass to derive a more precise 
mass accretion rate (knowing only one of the two is most often good enough). 

These conditions ensure that the failure of the
standard disk model is not due to a wrong reddening
(or reddening law), an unknown WD mass or distance.  
Although it has already been shown statistically the that the
failure of the standard disk model is due to the model itself, 
see \citet{pue07}.

\begin{deluxetable}{llll}
\tablewidth{0pt}
\tablecaption{System Parameters}
\tablehead{
Parameter                 & MV Lyr                   & BZ Cam                &   V592 Cas   \\ 
}
\startdata
$M_{\rm wd} <M_{\odot}>$  & 0.73-0.8$^{(1,2)}$       & 0.4-0.7$^{(7,8)}$     & 0.75$^{(13)}$     \\ 
$M_{2nd}    <M_{\odot}>$  & 0.3$^{(1)}$              & 0.3-0.4$^{(7)}$       & 0.21$^{(13)}$     \\ 
$i          <$deg$>$      & 10$\pm$3$^{(3,4,5)}$     & $\sim 15 < 40^{(9)}$ & 28$\pm 10^{(14)}$ \\ 
E(B-V)                    & 0$^{(6)}$                & 0.05$^{(10)}$         & 0.22$^{(13)}$     \\ 
 $P           <$hr$>$     & 3.19$^{(4)}$             & 3.69$^{(11,12)}$      & 2.76$^{(15)}$     \\ 
 $d           <$pc$>$     & $473\pm 37^{(1,2)}$      & 830$\pm$160$^{(9)}$  & 330-364$^{(13,14)}$ \\
\hline            
\enddata  
\tablecomments{References:   (1) \citet{hoa04};    (2) \citet{god12}  (3) \citet{sch81};
(4) \citet{ski95}; (5) \citet{lin05}; (6) \citet{bru94}; (7) \citet{luh85};  
(8) \citet{gre01};  (9) \citet{rin98};  (10) \citet{pri00}; (11) \citet{pat96}; 
(12) \citet{hon13}; (13) \citet{hoa09}; (14) \citet{hub98}; (15) \citet{tay98}}  
\end{deluxetable}

We list the system parameters in Table 1, 
from which it appears that MV Lyr is possibly the best choice:
it has no reddening, a very small inclination, and 
the distance, mass (and radius) and temperature of the WD
have all been well determined. BZ Cam too is a very good choice,
but its WD mass is not known accurately. BZ Cam is 
known to exhibit an {\it IUE} spectrum that has a remarkably flat 
(see section 2.2) continuum slope, therefore presenting a 
challenging test to our 
modified disk model. As for V592 Cas, it has a large
reddening value, but its parameters are otherwise relatively well known; 
its spectrum presents a UV continuum slope that is rather common for
disk-dominated CVs (as discussed in Sec.6).

In the following subsections, we describe in detail these three 
systems and their archival UV spectra, providing also their 
important characteristics.

\subsection{MV Lyrae}

MV Lyr has an  inclination $i\sim 10^{\circ} $ \citep{sch81,ski95,lin05}, 
and zero reddening \citep{bru94}. 
It is a VY Scl novalike system, 
i.e. spending most of its time in a high state 
and occasionally undergoing short-duration drops in brightness, 
or ``low state'', when accretion is almost shut off and the WD is revealed in the UV. 
It has a distance $d \sim 500 $pc and a WD mass of 
$M_{\rm wd} \sim 0.73-0.80 M_{\odot}$ \citep{hoa04,lin05,god12}.  
Compared to other CVs, the values of these parameters are relatively well known.  
The orbital period of the binary is 
$P=3.19$hr and the mass ratio was found to be 
$q = M_{2nd}/M_{\rm wd} =0.4$ \citep{sch81,ski95}. 
The system parameters are listed in Table 1.

MV Lyr spends most of its time in the high state, at a visual magnitude
of $\sim 12-13$, when the emission is primarily from the accretion disk.  
From time to time it drops into a low state, reaching a magnitude of
$\sim 17.5$, when the emission is possibly entirely from the WD. 
MV Lyr also has periods during which it is 
in an intermediate state at a magnitude of $\sim 14.-15$.  
MV Lyr was observed with {\it FUSE} and {\it IUE}, in a low state,  
with {\it IUE} in an intermediate state, 
and with {\it FUSE} and HST/STIS in a high state. 
The {\it IUE} observations of MV Lyr during its low state 
revealed
a hot WD possibly reaching 50,000K \citep{szk82} or higher \citep{chi82}. 
These results were later confirmed with a {\it FUSE} spectroscopic analysis 
in a low state \citep{hoa04}, 
during which the mass accretion rate was estimated to be no more than               
$\dot{M} \approx 3 \times 10^{-13}M_{\odot}$/yr and the WD temperature 
was found to be 47,000K. 
A follow-up analysis to study the 
different states of MV Lyr 
was carried out by 
\citet{lin05} who found that standard disk models did not fit the observed spectra.
Their models improved with the truncation of the inner disk, an isothermal
radial temperature profile in the outer disk, or both, and 
gave a mass accretion rate of the order of 
$3 \times 10^{-9}M_{\odot}$/yr for the high state and  
$1 \times 10^{-9}M_{\odot}$/yr for the intermediate state. 

The departure from the standard
disk model is more pronounced in the intermediate state
and most evident in the longer wavelengths of {\it IUE}.  
Since the high state was only observed down to $\sim 1,800$\AA\ , 
in our analysis we only model the system in the low and intermediate states 
extending above 3,000\AA\ . The modeling of the low state is first 
carried out (see section 5.1) to demonstrate the numerical method
and re-derive the system parameters.  
The observation log is recapitulated in Table 2.
The spectra are shown in Fig.1 together with a standard 
disk model for comparison. The shallow slope of the 
{\it IUE} spectra in the intermediate state appears clearly.

All of the {\it FUSE} spectra in this work were calibrated with the latest 
and final version of CalFUSE \citep{dix07}, and processed using our 
suite of FORTRAN programs, unix scripts, and IRAF procedures 
written for this purpose. 
The details can be found in \citet{god12}.

\begin{deluxetable*}{cccccrc}
\tablewidth{0pt}
\tablecaption{Observation Log}
\tablehead{
System & Telescope/ & Obs ID & Date (UT)  & Time (UT)& Exp. Time & State    \\ 
Name   & Instrument &        & yyyy-mm-dd & hh:mm:ss & seconds   & l-i-h                   
}
\startdata
MV Lyr  & FUSE     & C0410301 & 2002-07-07 & 11:56:39 & 11209 & l \\  
        & IUE/SWP     & 07296 & 1979-12-02 & 20:09:28 & 5400 & i  \\ 
        & IUE/LWP     & 06288 & 1979-12-02 & 18:52:24 & 4200 & i   \\  
        & IUE/SWP     & 10905 & 1980-12-27 & 11:29:05 & 2400 & l   \\ 
        & IUE/SWP     & 10906 & 1980-12-27 & 13:18:44 & 9000 & l  \\ 
        & IUE/LWP     & 09589 & 1980-12-27 & 12:12:51 & 3600 & l   \\  
        & IUE/LWP     & 09590 & 1980-12-27 & 15:50:09 & 7080 & l   \\  
BZ Cam  &  FUSE     & D9050101 & 2003-12-01 & 17:39:28 & 13187 & h\\  
        &  IUE/LWP  & 14485     & 1988-11-19 & 14:25:08 & 1500 & h \\   
        & IUE/LWP  & 14486     & 1988-11-19 & 14:45:36 & 1500 & h \\  
        & IUE/LWP  & 14487     & 1988-11-19 & 17:51:15 & 1500 & h \\  
        & IUE/LWP  & 14491     & 1988-11-20 & 13:44:17 & 1500 & h \\  
        & IUE/LWP  & 14492     & 1988-11-20 & 15:38:36 & 1380 & h \\  
        & IUE/LWP  & 14493     & 1988-11-20 & 17:31:46 & 1380 & h \\  
        & IUE/SWP  & 34775     & 1988-11-19 & 14:04:30 & 1800 & h   \\ 
        & IUE/SWP  & 34776     & 1988-11-19 & 16:23:32 & 3600 & h   \\ 
        & IUE/SWP  & 34777     & 1988-11-19 & 18:25:06 & 1440 & h   \\ 
        & IUE/SWP  & 34784     & 1988-11-20 & 12:16:20 & 3600 & h   \\ 
        & IUE/SWP  & 34785     & 1988-11-20 & 14:21:57 & 3600 & h   \\ 
        & IUE/SWP  & 34786     & 1988-11-20 & 16:16:59 & 3600 & h   \\ 
        & IUE/SWP  & 34787     & 1988-11-20 & 17:05:03 & 2520 & h   \\ 
V592 Cas & FUSE     & D1140101 & 2003-08-05 & 21:22:24 & 23751 & h \\ 
        & IUE/LWP & 12084      & 1981-12-05 & 19:48:23 & 1200 & h \\  
        & IUE/LWP & 12085      & 1981-12-05 & 22:24:02 & 1680 & h \\  
        & IUE/SWP & 15658      & 1981-12-05 & 18:48:15 &  900 & h \\  
        & IUE/SWP & 15659      & 1981-12-05 & 21:20:02 & 3600 & h \\  
\enddata 
\tablecomments{Note: (1) The flux level is h=high, i=intermediate, l=low. 
} 
\end{deluxetable*}

\begin{figure}[h]
\vspace{-5.cm}
\plotone{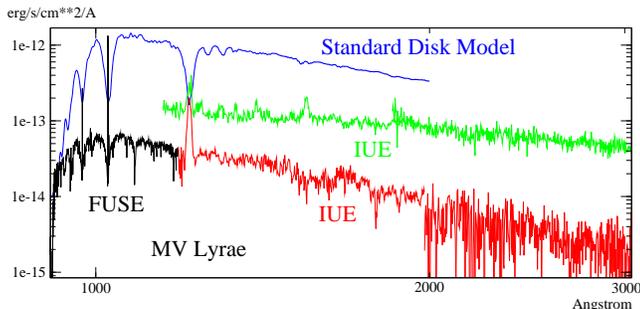}
\caption{
MV Lyrae has been observed in low, intermediate, and high states;
however, only the low and intermediate states have IUE spectral  
coverage all the way to 3200\AA. 
Shown here are the {\it FUSE} (black) and IUE (SWP+LWP) spectra of 
MV Lyr obtained in the low state (red), as well as an IUE (SWP+LWP) 
spectrum taken in an intermediate state (green).
The low state reveals the WD, while the intermediate state 
is expected to be dominated by emission from the disk.
However, the slope of the continuum of the spectrum obtained in 
the intermediate state is significantly shallower than the 
spectrum of a standard disk model, shown here in blue. 
The disk model has a central WD mass of $0.8M_{\odot}$, mass accretion rate of
$10^{-8}M_{\odot}$/yr and inclination of $41^{\circ}$. The
disk spectrum has been shifted upwards for clarity as it is 
displayed only to show the slope of its continuum (there is no
attempt here to fit the observed spectra with the theoretical disk spectrum).
The observed spectra have not been shifted.  
}
\end{figure}

\vspace{1.cm}

\subsection{BZ Camelopardalis}   

BZ Cam is a VY Scl NL type \citep{gar88,gre01}, 
with a relatively small reddening E(B-V)=0.05 \citep{pri00}, 
an orbital period of 221min \citep{pat96}, 
a distance of $830  \pm 160$pc \citep{rin98},  
and a moderately low inclination $\sim 12 < 40^{\circ}$ \citep{rin98}.  
Its WD mass is unknown but in our analysis we assume 
$ M_{\rm wd} \approx 0.4 M_{\odot} \le M_{\rm wd} \le 0.7M_{\odot}$  
\citep{gre01}. 
BZ Cam is not a typical CV in that it is surrounded by a faint
emission bow-shock nebula \citep{ell84,kra87,gre01},  
which is proof of an outflow \citep{pat96}. 
All the system parameters of BZ Cam are listed in Table 1 with 
their references.   

BZ Cam spends most of its time in a high state at a magnitude 
of $\sim 12-13$. It has been shown to drop occasionally to a magnitude
of $\sim 14$. All the UV spectra of BZ Cam were collected in 
the high state, but show some small variations in the continuum
flux level.  
The first Ultraviolet (UV) spectra of BZ Cam were obtained with 
{\it IUE} \citep{kra87,woo90,woo92,gri95}. 
We retrieved here  
13 {\it IUE} spectra from a single epoch (1988-11-19 \& 
1988-11-20) with the same continuum flux level. 
They consist of 6 {\it IUE} LWP spectra and 7 {\it IUE} SWP spectra, which 
we co-added and then combined (see Table 2). 
The {\it IUE} (SWP \& LWP) and optical continuum is consistent with a
$\sim$12,500K Kurucz LTE model atmosphere with $\log{g}=2.5$
\citep{pri00}, but it cannot
be attributed to the WD, because of its small emitting surface 
area. The {\it IUE} spectra of BZ Cam has a relatively shallow slope
when compared to an accretion disk. 
BZ Cam also has a {\it FUSE} spectrum \citep{fro12}, which was never   
previously modeled with realistic disk or WD atmosphere model spectra.   
The {\it FUSE} spectrum is affected by interstellar (ISM) absorption, 
mainly molecular hydrogen. 
The {\it FUSE} and {\it IUE} spectra of BZ Cam are shown (using a coarse binning
for clarity) in Fig.2. together with a standard disk model and a 
12,500K stellar atmosphere (with $\log{g}=8$) spectrum for comparison.  
The continuum flux level in the {\it FUSE} spectrum is slightly         
lower than the continuum flux level of the combined {\it IUE} spectrum
in the region where the two spectra overlap. This difference is likely due to
a small variation in the flux of BZ Cam, although we cannot exclude the
possibility that it could also be due, at least in part,
to the fact that the data were obtained
with different telescopes, gratings, detectors, and software.

\begin{figure}[h] 
\vspace{-2.cm} 
\plotone{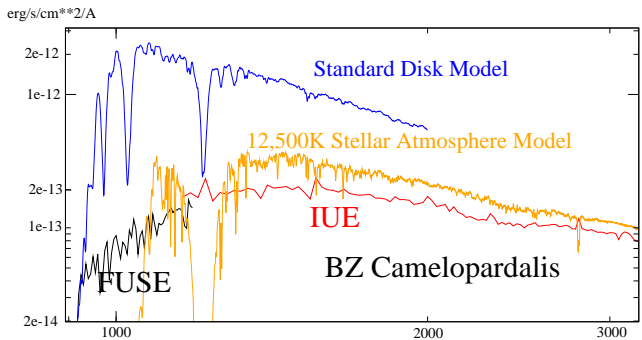} 
\caption{ 
The dereddened {\it FUSE} (in black) and  {\it IUE} (red) spectra of BZ Cam  
in its high state  
have been drawn on a log-log scale
together with the theoretical spectra of a standard disk model 
(blue)  
and a 12,500K stellar atmosphere (orange) for {\it comparison}.  
The accretion disk has a WD mass of $0.8M_{\odot}$, mass accretion
rate of $10^{-8}M_{\odot}$/yr and inclination $i=18^{\circ}$. 
The theoretical spectra have been shifted upwards for clarity. 
We do not attempt here to "fit" the theoretical spectra to the
observed ones.  
The relatively shallow slope of the {\it IUE} spectrum does not match 
the steeper slope of a  standard accretion disk, and is even more 
shallow than the 12,500K isothermal component.  
On this log-scaled figure, it appears that the {\it IUE} spectrum 
does not especially match the {\it FUSE} spectrum both in its continuum flux level 
and slope.  The observed spectra were not obtained at the same time, 
and since BZ Cam is a variable UV source,
the spectra cannot be combined together for modeling. 
In this figure, the {\it FUSE} and {\it IUE} spectra have been binned coarsely 
for clarity.
} 
\end{figure}

\subsection{V592 Cassiopeiae}

V592 Cas is an UX UMa subtype of novalike and has never been observed 
in a low brightness state. It has an inclination of $i=28^{\circ}$,  
a mass ratio of 0.19 \citep{hub98}, 
and its orbital period  is $P_{\rm orb}=$ 2.76 hr \citep{tay98,wit03}. 
It shows evidence for a bipolar wind outflow \citep{wit03,kaf09,pri04} 
with velocities reaching up to $v_{\rm wind} \sim 5000$ km$~$s$^{-1}$
\citep{kaf09}.  
The reddening toward the system is E(B-V)=0.22 \citep{car89}, 
and its distance is about 240-360 pc \citep{tay98}. 
The mass accretion rate of the system is believed to be      
of the order of  $\approx 1  \times 10^{-8} M_{\odot}$/yr
\citep{tay98,hoa09}.  
In the present work, we assume the WD to have a standard CV WD mass of 
$\sim 0.8M_{\odot}$ and a temperature of 45,000K \citep{hoa09}.

V592 Cas was observed with {\it IUE} on December 5th 1981. We retrieved the two
LWR spectra and two of the four SWP spectra that were obtained, combined  
and co-added  them to generate a spectrum with a spectral range 
from $\sim 1150$ \AA\ to $\sim 3,200$ \AA\ .  
With a flux of a few $10^{-13}$erg/s/cm$^2$/\AA , 
the spectrum has a relatively good signal in the SWP segment but it is
rather noisy in the LWR segment, as shown in Fig.3 (in red). 
V592 Cas has 3 {\it FUSE} data sets, obtained on August 5, 7 and 8, 2003. 
The data sets have all the same quality and about the same exposure time
and reveal the same spectrum. 
We decided to retrieve the first data set (D1140101) and extracted  
a final spectrum following the same procedure we used for BZ Cam
and MV Lyr. 
The {\it FUSE} spectrum V592 Cas has the same continuum
flux level as the {\it IUE} spectrum in the region where the two spectra overlap. 
Like the {\it FUSE} spectrum of BZ Cam, the {\it FUSE} spectrum of V592 Cas 
is heavily affected by ISM molecular hydrogen lines, appearing in 
Fig.3 (in black) as if the spectrum had litterally been sliced. 

In Fig.3 we also plot two standard disk models for comparison.  
The {\it FUSE} continuum is in better agreement with 
the $10^{-8}M_{\odot}$/yr model, while the {\it IUE} continuum slope
is even shallower than the $10^{-9}M_{\odot}$/yr model. 
These indicates  that the combined {\it FUSE} + {\it IUE} spectrum does 
not agree very well with the standard disk model. However, the discrepancy 
is not as strong as for MV Lyr and BZ Cam.

\begin{figure}
\vspace{-6.cm} 
\plotone{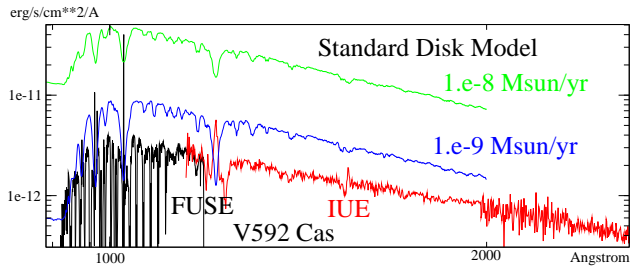}             
%\vspace{-3.cm} 
\caption{The dereddened {\it FUSE} (solid black) and {\it IUE} (red) 
spectra of V592 Cas in its normal high state have been plotted 
on a log-log scale against two standard disk model spectra for 
{\it comparison}. 
The theoretical spectra have been shifted upward for clarity, 
there is no attempt here to "fit" the observed spectra with
the theoretical ones. 
The standard disk models have a central WD mass of $0.8M_{\odot}$
with mass accretion rate of 
$\dot{M}=10^{-8}M_{\odot}$/yr (green/upper plot) 
and $\dot{M}=10^{-9}M_{\odot}$/yr (blue/lower plot),  
and inclination of $41^{\circ}$ and $18^{\circ}$
respectively. The {\it FUSE} flux continuum {\it shape} 
is in better agreement with the $10^{-8}M_{\odot}$/yr model 
at wavelengths shorter than 1,000\AA ,  
while the {\it IUE} flux continuum {\it slope} 
is even shallower than the $10^{-9}M_{\odot}$/yr model. 
These are clear indications that the spectra do not agree with the
standard disk model.
The {\it FUSE} spectrum is strongly affected by molecular hydrogen   
absorption lines from the interstellar medium (ISM).  
}
\end{figure}

\section{{\bf The Modified Disk Model}} 

We wish to model the three NL systems in high and intermediate states using a 
disk model generated by modifying the standard disk model. 
Therefore, we first review the standard disk model, and then we introduce
the modification we make. 

\subsection{The Standard Disk Model} 
 
In the standard disk theory \citep{pri81}, 
the total accretion luminosity that can be 
released by accretion onto a star is given by:  
\begin{equation} 
L_{\rm acc} = \frac{G M_{\rm wd} {\dot{M}}} {R_{\rm wd}}, 
\end{equation}
where $G$ is the gravitational constant, $\dot{M}$ the mass accretion
rate, $M_{\rm wd}$ the mass of the
accreting star (here a WD) and $R_{\rm wd}$ its radius. 

In disk systems, accretion energy is released through the disk.  
All the current disk models are based on the Shakura-Sunyaev
`alpha' disk model known as the {\it standard disk model}
\citep{sha73,lyn74}. 
The disk is assumed to be geometrically thin (in the vertical
dimension) and the energy dissipated between adjascent
rings of matter ($\Phi_{\nu}$, the dissipation function 
due to the shear) is radiated locally in the
vertical direction:  $\sigma T_{\rm eff}^4 = \Phi_{\nu}$. 
The viscosity $\nu$ is due to turbulence and has
been shown to come from a magneto-rotational instability 
\citep{bal91}.            
The turbulent viscosity
is parametrized with an {\it a priori} unknown parameter $\alpha$ 
($\alpha < 1$), which is the reason the standard disk model is 
also called the alpha disk model.  
The standard disk model is obtained from the  
steady-state Navier-Stokes equations in cylindrical coordinates 
$(R,\phi,z)$). The equations are reduce to one dimension ($R$) 
after integrating in the vertical dimension ($z$),
and assuming axi-symmetry ($\partial \phi / \partial = 0$).
Using the conservations of mass (for $\dot{M}$) 
and angular momentum, and imposing the boundary condition that  
the angular velocity gradient vanishes
($\partial \Omega / \partial R = 0$)  at the inner boundary $R=R_{\rm wd}$, 
one obtains the standard disk model \citet{pri81}. 
In that approximation the boundary layer between the star and disk 
is geometrically thin, its thickness can be neglected 
$\delta_{\rm BL} << R_{\rm wd}$.

\subsection{The New Boundary Condition} 
 
We modify the standard disk model 
by truncating the inner disk at a radius $R_0 > R_{\rm wd}$. 
In order to do this correctly, one cannot just remove the inner
disk between $R_{\rm wd}$ and $R_0$ from a standard disk model starting
at $R_{\rm wd}$, instead one has to generate 
a standard disk model starting at $R=R_0$. 
We give here below, several reasons for having a disk starting 
at $R_0$ rather than $R_{\rm wd}$.  

The optically thin boundary layer simulations of \citet{pop93}
and \citet{pop99} 
{\bf show}
 that the optically thin boundary layer
has a size $\delta_{\rm BL} \approx R_{\rm wd}$ and that the angular
velocity gradient vanishes at the outer edge of the boundary layer.
Namely, the no-shear boundary condition 
$\partial \Omega / \partial R = 0$ has to be imposed at 
$R=R_0 =R_{\rm wd} + \delta_{\rm BL} > R_{\rm wd}$. 

If the inner hole in the disk is due to a weakly magnetized WD truncating
the disk \citep{ham02} at $R_0$ (as in intermediate polars - IPs), 
one still has the same new boundary conditions, as $R_0$ is now  
the corotation radius, which by definition is the radius 
where $\partial \Omega / \partial R = 0$, since $\Omega =$constant there. 

If the inner hole in the disk is due to evaporation
\citep{ham00},            
the same new boundary conditions can also be imposed as the disk 
only starts at the outer radius (now $R_0$) of the evaporated inner disk.   

Also, it is important to note that the radius of the WD is expected to 
increase as the WD temperature increases. For example a 
10,000K WD of mass $0.7M_{\odot}$ has a radius 
$R_{\rm wd}=7.9 \times 10^8$cm, 
and at 50,000K the radius increases by 16\% to 
$R_{\rm wd}=9.2 \times 10^8$cm \citep{woo95}. 
The increase in radius is even more pronounced for
a smaller mass WD and basically reaches a factor of 2 for a $0.4M_{\odot}$ WD,  
with a zero-temperature radius of $R_{\rm wd}=1.1 \times 10^{9}$cm 
versus a 50,000K radius of $R_{\rm wd}=2 \times 10^{9}$cm \citep{woo95}. 
Since the VY Scl systems are notorious for having a temperature much hotter 
than other CVs at the same period, the change in 
radius cannot be ignored. So far, most standard disk model spectra were 
computed assuming a constant (zero-temperature) radius
\citep{wad98,god12}, thereby not taking into account the increased WD radius 
and possibly overestimating the disk temperature. Modeling the disk 
properly requires that the WD temperature be known, and requires a
fine tuning of the modeling such as the one performed in \citet{lin08}.

The modified disk model can therefore be applied
to different inner disk structures, including the magnetically truncated
disks in IP systems, and around hot CV WDs to provide a more realistic 
approach for the modeling of the disk.   

\subsection{The Temperature of the Modified Disk} 

Our modified disk model differs from the standard disk model in  
that we now impose the no-shear boundary condition at $r=R_0 > R_{\rm wd}$. 
Namely,  
\begin{equation} 
\frac{\partial \Omega}{\partial R} = 0,~~~~~~ R=R_0,  
\end{equation} 
and, following \citet{pri81}, we 
assume a Keplerian velocity $\Omega = \Omega_K$ 
for $R \ge R_0$.  

With these assumptions and boundary conditions one obtains  
the disk radial temperature profile 
\begin{equation} 
T_{\rm eff} (R) = \left\{ 
\frac{3G M_{\rm wd} \dot{M}}{8 \pi \sigma R^3} 
\left[ 1 - \sqrt{ \frac{R_0}{R} } \right] \right\}^{1/4} , 
\end{equation} 
where the angular velocity in the disk has been substituted with 
its Keplerian value. 
For convenience, this expression can be   written  as                           
\begin{equation} 
T_{\rm eff}(x) = T_0 x^{-3/4} ( 1 -x^{-1/2})^{1/4}, 
\end{equation} 
with
\begin{equation}  
T_0 = 64,800K \times 
\left[ \left( \frac{M_{\rm wd}}{1 M_{\odot}} \right)  
\left( \frac{\dot{M}}{10^{-9} M_{\odot}/yr} \right)  
\left( \frac{R_0}{10^9 cm } \right)^{-3} \right]^{1/4}  
\end{equation}  
where $ x=R/R_0 $.  
This model gives a maximum temperature ($T_{max}=0.488T_0$) at 
$x=1.36$, and $T=0K$ at $R=R_0$ ($x=1$).  
\\
\indent                
The expression for the temperature of the modified disk model 
is the same as for the standard disk model,  
except for the star radius $R_{\rm wd}$, which 
has simply been replaced by the inner disk radius $R_0$.  
For example, for an inner disk radius twice as large as the star radius
($R_0=2R_{\rm wd}$), the maximum 
effective surface temperature in the disk (reached at $R=1.36R_0$) has
dropped to $\sim$65\% the value it has in the standard disk model 
(at $R=1.36R_{\rm wd}$).  In Fig.4 we show the temperature profile for a modified disk
model for different value of the truncated radius $R_0$. 
The mass of the WD is $0.8M_{\odot}$, and the mass accretion rate is 
$1 \times 10^{-9}M_{\odot}$/yr. The radius $R_0$ is given in units of
the WD radius $R_{\rm wd}$ (which is itself obtained from the 
WD mass-radius relation, e.g. \citet{woo95}).  
From Fig.4, it is clearly apparent that truncating a standard disk model 
by solely removing the region $R< 2R_{\rm wd}$ 
gives a disk model with a much
higher temperature in the inner disk than generating a model starting 
at $R_0=2R_{\rm wd}$. As the size of the BL is possibly not negligible
and as the WD can easily increase its radius with increasing temperature,
the standard disk model can over-estimate the temperature in the inner disk 
by a large fraction. 
\\
\indent 
While our modified disk model is actually a standard disk model in 
which the radius of the star $R_{\rm wd}$ has been replaced by the 
inner radius of the disk $R_0$ (where $R_0>R_{\rm wd}$), we use
through this paper the terms {\it modified disk} and {\it standard
disk} to differentiate between the two.

\begin{figure}[h] 
\vspace{-3.5cm} 
\plotone{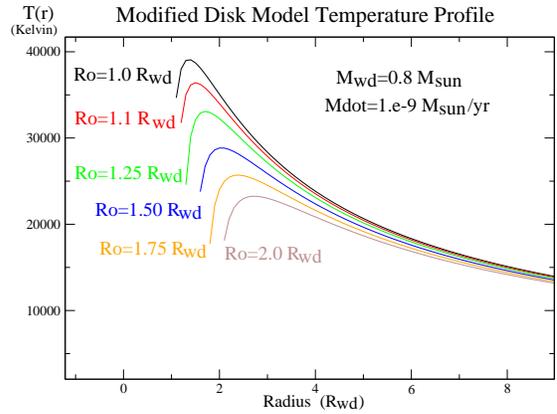}
\caption{ 
The temperature profile of the modified disk model is shown
as the radius $R_0$ (in units of $R_{\rm wd}$) 
of the inner boundary of the disk is increased from 
one stellar radius (the standard disk model) 
to two stellar radii. Here the WD mass is $0.8M_{\odot}$
and the mass accretion rate is $\dot{M}=1 \times 10^{-9}M_{\odot}$/yr. 
As $R_0$ increases, the disk temperature profile flatens and the 
disk luminosity decreases. For low $M_{\rm wd}$ and $\dot{M}$ and for large
$R_0$, the temperature becomes almost isothermal.  
}
\end{figure}

\subsection{The Luminosity of the Modified Disk} 

The disk luminosity $L_{\rm disk}$ is given by integrating $\Phi_{\nu}$ 
(or $\sigma T^4_{\rm eff}$) over the surface of the disk 
\begin{equation} 
L_{\rm disk} = 2 \int_{R_0}^{\infty} \Phi_{\nu} 2 \pi R dR = 
\frac{1}{2} \frac{G M_{\rm wd} \dot{M}}{R_0} . 
\end{equation}  
In the standard disk model the disk luminosity amounts 
to half the accretion energy, in the modified disk model it is
smaller than half the accretion energy. 
For example, for $R_0=2R_{\rm wd}$ the disk luminosity is only half the disk
luminosity expected for a standard non-truncated disk. 
{\bf The important point is that a truncated disk cannot be generated 
by simply truncating a standard disk model, but it has to be obtained
as a standard disk model in which $R_{\rm wd}$ is replaced by $R_0$, which produces
a much lower disk temperature and luminosity.}  
The idea of replacing $R_{\rm wd}$ by $R_0$ 
was discussed briefly in the modeling of IX Vel 
\citep{lon94} but was never followed through. Similarly the modeling of 
MV Lyr by Linnell et al. (2005) includes a truncated disk, but with 
the `usual` temperature profile of the standard disk model ($R_{\rm wd}$
was not replaced by $R_0$, instead the standard disk was just truncated).  
\subsection{The Boundary Layer Luminosity} 
The remaining fraction of the accretion energy is  
dissipated at the inner edge of the disk, where the disk meets
the surface of the star($r=R_{\rm wd}$), in the so-called boundary layer
(BL). 
The fast (Keplerian) rotating ($\Omega_K(R_0)$) matter
adjusts itself to the slowly rotating stellar surface ($\Omega_{\rm wd}$) and
dissipates its remaining kinetic energy.  The 
luminosity of the boundary layer $L_{\rm BL}$, in the standard disk model,  
is of the same order of magnitude as $L_{\rm acc}/2$, though usually smaller 
as some of the energy goes into spinning up the WD \citep{klu87}.       
This is true for optically thick boundary layers. 

X-ray observations of disk-dominated CVs            
(e.g. \citet{van96,bal14}) show 
that the BL is optically thin and that the X-ray luminosity 
is much smaller than the disk luminosity, the BL is underluminous. 
For the optically thin boundary layer, one can only refer to the
work of \citet{pop93,pop99} and to observations. 
\subsection{The Fate of the Remaining Accretion Energy} 
If the disk emits less than half the accretion energy, and the
boundary layer luminosity (in the X-ray) is itself only a fraction of the disk 
luminosity, one might ask where does the remaining energy go? 
When an accretion flow cannot cool efficiently, due to being optically thin, 
the energy is advected with the flow into the outer layer of the 
star \citep{abr95}. 
The advected BL energy will then increase the WD temperature,  
and will likely drive an ouflow as in Advection Dominated Accretion Flows - 
ADAF \citep{nar95}. 
This is supported by
three observational facts: (i) The WD temperature of NLs               
is rather elevated when compared to other CVs with the same period.    
(ii) Many NL systems exhibit hot, fast wind outflows 
seen in P Cygni line structure and blue-shifted 
absorption. These winds originate in the inner disk/BL region but their
source of energy has not yet been identified. 
Some CV systems even have visible ejected nebular 
material (e.g.  BZ Cam, \citet{ell84} ). 
(iii) Recent X-ray observations of NLs \citep{bal14}  
suggest that high-state NLs may have optically thin BLs merged with ADAF-like
flows and/or hot coronae. \citet{bal14} showed that the observed 
discrepancy between the X-ray/boundary layer and disk luminosity can 
be accounted for if the boundary layer energy is heating the WD 
through advection ($T_{\rm wd} \sim 50,000$K) and is also used to drive the wind
outflows. The efficiency of the boundary layer to radiate its energy is
then greatly reduced ($< 0.01$), consistent with the observed ratio of the 
X-ray to the disk UV-Optical luminosities.  

For this reason, we expect NLs to be modeled with the modified disk
model and the addition of a hot WD, itself possibly with an inflated  
radius.  

\subsection{The Outer Disk Modeling} 
In CVs the accreting WD is the more massive component 
and the Roche lobe radius is always larger than half the binary
separation, i.e. $>a/2$.  
Due to the impact of the stream (from the first Lagrange point L1) 
on the rim of the disk and due to 
tidal interaction, the outer disk is expected to have a temperature
larger than dictated by both the standard and modified disk models
\citep{bua01}.                     
We assume here that the outer region of the disk  
has a temperature of 10,000K to 12,000K.  
We note that such a practice is not uncommon, and some authors
(e.g. Linnell et al. 2005) have considered disk models with an
elevated temperature ($\sim 10-12,000$K) in the outer region of the disk.

\section{{\bf UV SPECTRAL MODELING}}   
              
\subsection{TLUSTY and SYNSPEC Codes}
              
We use the FORTRAN suite of codes TLUSTY, SYNSPEC, ROTIN and 
DISKSYN \citep{hub88,hub94,hub95}  
to generate synthetic spectra of stellar atmospheres and disks.
The codes include the treatment
of hydrogen quasi-molecular satellite lines (low temperature) and
LTE and NLTE options (high temperature). SYNSPEC generates a continuum
with absorption lines. In the present case we do not generate emission lines. 
For disk spectra we use solar abundances 
and for stellar spectra we vary the abundances
as needed. 
An introductory guide, a reference manual and an operational manual
to TLUSTY and SYNSPEC have just been released and are available for 
full details of the codes \citep{hub17a,hub17b,hub17c}.

\subsection{WD stellar atmosphere models} 
              
The code TLUSTY is first run to generate a one-dimensional (vertical)
stellar atmosphere structure for a given surface gravity,
effective temperature and surface composition of the star. 
Computing a single model is an iterative process that needs to converge.
In the present case, we treat hydrogen and helium explicitly, 
and and treat nitrogen, carbon and oxygen implicitly 
\citep{hub95}. 
                
The code SYNSPEC is then run, 
using the output stellar atmosphere model from TLUSTY
as an input, and generates a synthetic stellar spectrum over a given
(input) wavelength range (it has capabilities to cover a spectral
range from below 900\AA\  and into the optical).   
The code SYNSPEC derives the detailed radiation and flux distribution of 
continuum and lines and generates the output spectrum \citep{hub95}.           
SYNSPEC has its own chemical abundances input to generate lines for 
the chosen species.  
For temperatures above 35,000K the approximate NLTE treament of lines 
is turned on in SYNSPEC. 
   
Rotational and instrumental broadening as well as limb darkening 
are then reproduced using the routine ROTIN. 
In this manner we have already generated a grid of WD synthetic
spectra covering a wide range of temperatures and gravities 
with solar composition (see http://synspecat.weebly.com/ ). 
    
\subsection{Disk Models} 
    
The disk spectra are generated by dividing the disk into rings, 
each  with a given radius $r_i$ and temperature ($T(r_i)$), density and effective
vertical gravity obtained from the modified disk model
for a given stellar mass, inner and outer disk radii, 
and mass accretion rate.  
      
The code TLUSTY generates a one-dimensional 
vertical structure for each disk ring.  
The input parameters are the mass accretion rate,
the mass of the accreting star, the radius of the star, and the radius of the 
ring. The radius of the star in our modified disk model
is set to be the inner radius of the disk $R_0$.  
   
SYNSPEC is then used to create a spectrum for each ring,
and the resulting ring spectra are integrated into a disk spectrum 
using the code DISKSYN, 
which includes the effects of (Keplerian) rotational broadening,
inclination, and limb darkening.

For disk rings below 10,000 K, we 
use Kurucz stellar spectra of appropriate temperature and surface
gravity.
   
In Fig.5 we present the synthetic spectra of two disks generated with 
the suite of codes: a standard disk 
model (in black) and a modified disk model (in red). 
Both models have a $0.55 M_{\odot}$ WD accreting at a rate 
of $\dot{M}=10^{-9}M_{\odot}$/yr, and are displayed for an inclination
$i=40^{\circ}$. The standard disk has an inner radius
at $R_0=R_{\rm wd}$, while the modified disk has an inner radius   
at $R_0=2R_{\rm wd}$. The slope of the continuum of the modified disk 
is almost flat.

\begin{figure}[h]
\vspace{-2.cm}
\plotone{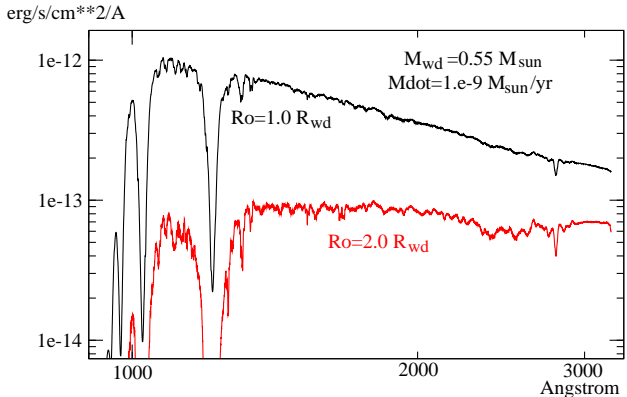}
\vspace{-2.cm}
\caption{
The synthetic spectra of two accretion disks are shown in the 
ultraviolet range. Both disk models have a $0.55 M_{\odot}$ 
WD accreting at a rate of $10^{-9}M_{\odot}$/yr. 
The upper spectrum (in black) is a standard disk model with the  inner disk 
radius at $R_0=R_{\rm wd}$, the lower spectrum (in red) is that of a 
modified disk with the inner radius at $R_0=2 R_{\rm wd}$. Because of the 
change in the disk temperature profile, the lower spectrum is characterized  
by a continuum slope that appears almost flat on this log-log scale. 
The continuum exhibits an ``elbow'' at $\lambda \sim 2400$\AA , which is 
a real feature observed in the {\it IUE} spectrum of BZ Cam.      
}
\end{figure}

\section{{\bf Results}}

\subsection{MV Lyrae}

\paragraph{The Low State} 

The modeling of the {\it FUSE} spectrum of MV Lyr in the low state was 
presented elsewhere \citep{god12} and gives identical results
to the modeling of the combined {\it FUSE} + {\it IUE} spectrum presented here,
which covers a much larger wavelength range. 
However, it is important to re-derive and confirm 
the distance to MV Lyr, its WD mass, radius, 
temperature and rotational velocity, 
because these are important parameters serving as input 
for the modeling of the intermediate state. Namely,  
these parameters are used as  constrains in the modeling of the 
combined {\it FUSE} + {\it IUE} spectrum of MV Lyr in the intermediate state.

While the {\it FUSE} spectrum presents many absorption lines 
commonly detected  in the atmosphere of accreting WDs
(see \citet{god12} for details), the {\it IUE} (SWP+LWP) spectrum 
is more noisy and presents only a continuum, in which 
the only identified absorption feature is the C\,{\sc iv} 1550 line. 
The {\it FUSE} and {\it IUE} spectra match in their flux 
level and slope, in spite of the fact that they were obtained with different
telescopes more than 20 years apart. This is possibly a sign that MV Lyr 
had reached a similar (if not identical) low state in which the WD completely
dominates the FUV \citep{hoa04}.

Our fitting results yield \citep{god12}:    
a WD temperature $T_{\rm wd}=44-47,000$ K,   
with a projected stellar rotational velocity $V_{rot} \sin{i} = 150-250$km/s,  
chemical abundances $Z = 0.1-1.0  Z_{\odot}$, 
and stellar surface gravity  $\log{g}=8.06-8.28$ 
($M_{\rm wd}=0.7-0.8M_{\odot}$ with $R_{\rm wd}$=9,010-7,450km).  
The resulting distance we obtained is $\sim$464pc for a 
$0.8M_{\odot}$ WD and $\sim 560pc$ for a $0.7M_{\odot}$ WD. 
In Fig.6 we present a 45,000K WD fit as one of the solutions 
we obtained. 
The remarkable characteristic of this model fit is that, although it was 
originally generated to fit the {\it FUSE} spectrum alone \citep{god12},
as shown in Fig.6, 
it actually also fits the {\it IUE} spectrum all way to 
the longest wavelengths $\sim$3000\AA\  (see Fig.7; 
the region beyond 3000\AA\  is very noisy and has been discarded).   
From this fit one obtains relatively accurate values for 
some of the important system parameters.  
  
These values of the parameters 
are consistent with the analysis of \citet{hoa04} who obtained
$T_{\rm wd}=47,000$K with $\log{g}=8.25$ for a non-rotating WD, and  
$T_{\rm wd}=44,000$K with $\log{g}=8.22$ for a WD rotating at 200km/s with
elemental abundances of C=0.5, N=0.5 and Si=0.2 (solar),
and a derived distance of $505 \pm 50$pc.

\begin{figure}[h]
\vspace{-2.cm} 
\plotone{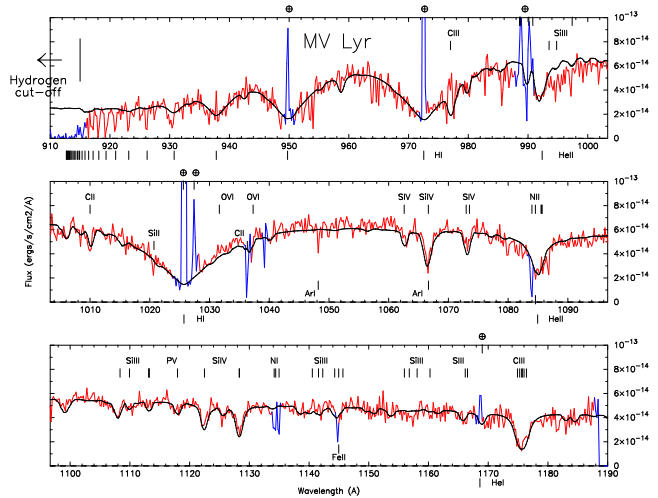}
\caption{
A synthetic WD atmosphere spectrum fit (in black) to the combined 
{\it FUSE} + {\it IUE} 
spectrum of MV Lyr (in red) in the low state (the blue shows the
contamination by terrestrial airglow). This figure shows the
{\it FUSE} range, while the next figures shows the {\it IUE} range.  
The sharp emission lines are artifacts due to air glow (marked with a  
plus sign inside a circle at the top of each panel) or indirect reflection of
sunlight inside the telescope (such as e.g. the  He\,{\sc ii} 1168\AA\ ).   
The WD model has a temperature of $T_{\rm wd}=45,000$K, 
solar composition and a projected rotational  velocity 
of $V_{rot} \sin(i) =200$km/s. 
} 
\end{figure}

\begin{figure}[h]
\vspace{-1.cm} 
\plotone{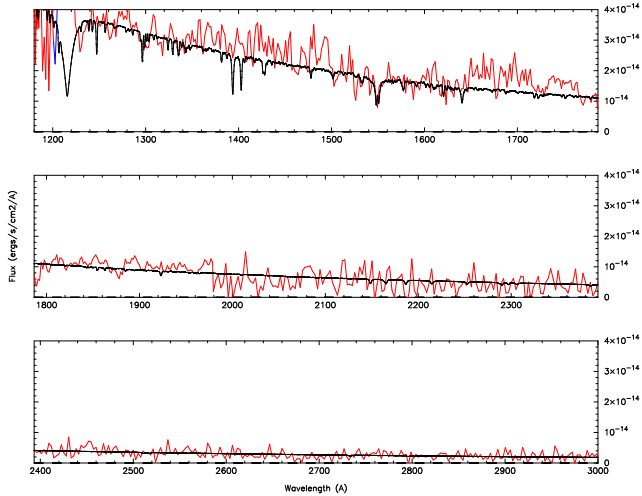}
\caption{
The synthetic WD atmosphere spectrum fit (in black) to the combined 
{\it FUSE} + {\it IUE} spectrum (in red) of MV Lyr in the low state is presented here
showing only the {\it IUE} range.  
Note that the vertical scale is different than in Fig.6 for clarity.  
} 
\end{figure}

\paragraph{The Intermediate State}

MV Lyr was caught in an intermediate state with {\it IUE}, 
covering that region of the spectrum where the continuum
slope departs strongly from a standard disk model all the way
to $\sim $3,200\AA\ . In order to model the intermediate state spectrum, 
we use the parameters we derived from fitting the low state:  
we adopt a WD mass of $M_{\rm wd}=0.73M_{\odot}$, with a 
temperature of $\sim$45,000K, a radius $R_{\rm wd} =$ 8,500km 
and  a distance of $\sim$500pc. 
We use the system inclination $i=10^{\circ}$ from the litterature
(see Table 1). 

We first attempt to fit the combined {\it IUE} SWP+LWP spectrum with 
a standard disk model, where the only varying parameter is 
the mass accretion rate (since the inclination and WD mass are
fixed). The inner disk radius is simply $R_{\rm wd}$ and the 
outer radius of the disk is place at a distance  
$a/3$, where $a$ is the binary separation. Using the system parameters
we have $a/3 \sim 30R_{\rm wd}$.  
We also include the expected contribution of the 45,000K WD.
In order to fit the distance to the system (by scaling the theoretical
spectrum to the observed one), we find a mass accretion rate of  
$8 \times 10^{-10}M_{\odot}$/yr. However, this model does not fit 
the observed spectrum at all (see Fig.8), 
the theoretical spectrum is too ``blue'' when 
compared to the observed spectrum, consistent with the statistical
analysis of \citet{pue07}.

Next, we fit the observed spectrum with modified standard disk models.  
For each given mass accretion rate we built disks with an increasing 
larger inner radius $R_0/R_{\rm wd}=$ 1.0, 1.2, 1.5, 1.7, 2.0, 2.5
and 3.0. 
We kept the outer disk radius at $30R_{\rm wd}$ and set the outer disk 
at a temperature around 10, 11 or 12,000K. The width of this isothermal
outer disk region was also varied from $1 R_{\rm wd}$ to $10 R_{\rm wd}$
(i.e. to a maximum of 1/3 of the disk radial extent).
The increase in the inner disk radius, produces a decrease of flux 
more pronounced in the shorter wavelength region of the spectrum, while 
the setting of the outer isothermal disk region produces a relative
increase of flux in the longer wavelength region.  
We find that a best fit is obtained for a mass
accretion rate of $2.4 \times 10^{-9}M_{\odot}$/yr with an inner hole 
of size $2.0R_{\rm wd}$ and an outer disk, from $20R_{\rm wd}$ to 
$30R_{\rm wd}$, set with a temperature of 10,000K.  
This model is presented in Fig.9
and also includes the contribution of 45,000K WD.  
The strong emission lines are shown in blue, there are more emission
lines on the edges of the Ly$\alpha$ profile which 
we are not modeling, as they form in an optically thin medium.   
The fit to the continuum 
is much improved, when compared to the standard disk model.

\begin{figure}[h] 
\vspace{-7.cm} 
\plotone{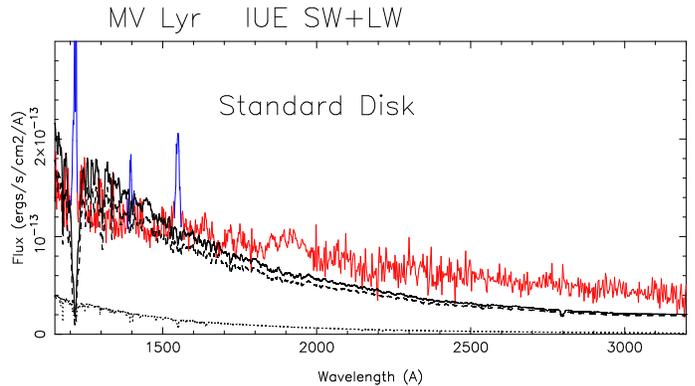}
\caption{ 
The combined SWP + LWP {\it IUE} spectrum of MV Lyra in an intermediate state 
(solid red line; strong emission lines in blue)  
is modeled with a hot WD (45,000K)  
and a standard accretion disk  
with $\dot{M}=8 \times 10^{-10}M_{\odot}$/yr and $i=10^{\circ}$.
The inner radius of the disk is at $1R_{\rm wd}$ and the outer
radius is at $30 R_{\rm wd}$.  
The combined model WD+disk is the solid black line,  
the dotted line in the lower part of the panel shows the WD contribution,
the dashed line shows the contribution from the disk.
The model cannot match the continuum flux level, it is far too `blue`.
}
\end{figure}

\begin{figure}[h] 
\vspace{-5.cm} 
\plotone{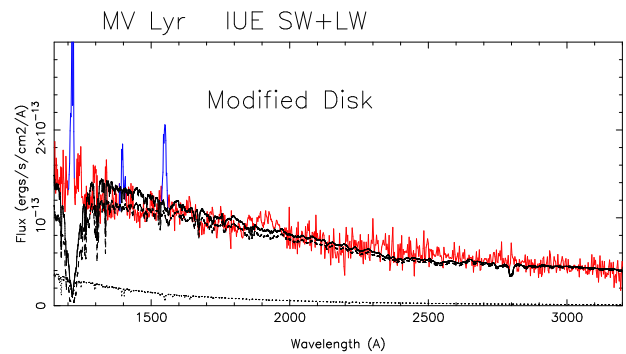} 
\caption{
The same {\it IUE} spectrum (in red) 
as in Fig.8 is modeled here with a hot WD (dotted line) and our  
modified disk model (dashed line); the combined model is the 
solid black line. 
The mass accretion rate is  $\dot{M}=2.4 \times 10^{-9}M_{\odot}$/yr 
and $i=10^{\circ}$.
The inner edge of the disk is placed at $R_0=2.0R_{\rm wd}$. 
The outer region ($20R_{\rm wd}$ to $30R_{\rm wd}$) of the disk has 
been set at a temperature of 10,000K. 
The strong emission lines are shown in blue, there are more emission
lines on the edges of the Ly$\alpha$ profile which 
we are not modeling, as they form in an optically thin medium.   
} 
\end{figure}

\subsection{BZ Camelopardalis}  

BZ Cam has a {\it FUSE} spectrum and an {\it IUE} spectrum that do not have
exactly the same flux level, in addition the {\it IUE} spectrum is consitent
with a rather cold component, while the {\it FUSE} spectrum seems to be consistent
with a hotter component. Consequently, we first model these two 
spectra separately, and next we try to model the two spectra in 
a self-consistent manner.

\paragraph{Modeling the {\it FUSE} Spectrum} 

The {\it FUSE} spectrum shows the presence of a hot component contributing flux
all the way to the shortest wavelengths of {\it FUSE}. 
We try a standard disk model, assuming $M_{\rm wd}=0.8M_{\odot}$, 
and we find that the mass 
accretion rate has to be as large as $10^{-8}M_{\odot}$/yr 
to produce the observed flux in the short wavelengths of {\it FUSE}. 
This model, however, gives a distance of more than 2kpc, i.e.
more than twice the accepted distance to the system. 
As we decrease the WD mass to $0.55M_{\odot}$, the distance 
decreases slghtly to 1.8kpc, 
and for a $0.35M_{\odot}$ WD the distance is still too large,
however, the flux 
in the shorter wavelengths of the theoretical spectrum is 
now too low.  
For a slightly lower mass accretion rate of $10^{-8.5}M_{\odot}$/yr,
only the larger WD model with $M_{\rm wd}=0.8M_{\odot}$ provides enough
flux in the short wavelengths, but here too the distance is too large
and reaches 1.4kpc. 

These results show 
that the single disk model does not provide a satisfactory fit. 
Since our modified disk model provides colder disks, we do not
try to use such a modeling for the {\it FUSE} spectrum, 
as it would not provide enough flux in the shorter wavelengths.

Instead, we try single WD models. 
The best fit WD model has $M_{\rm wd}=0.4M_{\odot}$,  
$T_{\rm wd}=45,000$K, solar composition and $V_{rot} \sin{(i)} =200$km/s, 
and the distance obtained from the fit is 791pc, 
when the WD radius is set to $R_{\rm wd}=18,540$km.  
This model is shown in Fig.10,   and 
includes a basic modeling of the 
ISM molecular hydrogen absorption lines (see \citet{god07} for details). 
In the short wavelengths ($\sim$960\AA\ upper panel) 
the model has too much flux, while  
in the longer wavelengths (lower panel) the model has too little flux. 
A 40,000K WD gives a distance of 
654pc while a 50,000K WD gives a distance of 933pc. 
We note that at a temperature of about 50,000 K the radius of the 
$0.4M_{\odot}$ WD model ($\sim 2 \times 10^{9}$cm) 
is about twice that of the zero-temperature model 
($\sim 1 \times 10^{9}$cm).  

Next, we increase the WD mass to $0.7M_{\odot}$, 
and accordingly decrease the radius to $\sim 0.9 \times 10^{9}$cm, 
we find that the scaled distance    
from the fit becomes too short for the same temperature, and 
higher temperature models ($T>50,000$K) do not provide the best fit. 
The fit of the 45,000K $0.7M_{\odot}$ WD 
is as good as the $0.4M_{\odot}$ fit with the same temperature, 
and indicates that a $0.7M_{\odot}$ WD  
with a radius of $\sim 2 \times 10^{9}$cm provides a valid solution
indicating that maybe the radius of the WD is inflated.   

We also try to add a standard disk model to the WD disk model,
but we find that disk model deteriorates the fit and increases  
the distance well above 1kpc.  

To summarize, the {\it FUSE} spectrum of BZ Cam is best fitted with 
a single WD with a temperature of 45,000K $\pm 5,000$K, and a 
mass $0.4-0.7M_{\odot}$ (from the litterature) as long as the WD
radius is  $\sim 2 \times 10^{9}$cm (to fit the known distance to 
the system). Since this radius really corresponds to a $0.4M_{\odot}$
WD at $T\sim 50,000$ K (see section 3.2), 
this is the WD mass we adopt as the best solution.

\begin{figure}[h] 
%\vspace{-7.cm} 
\plotone{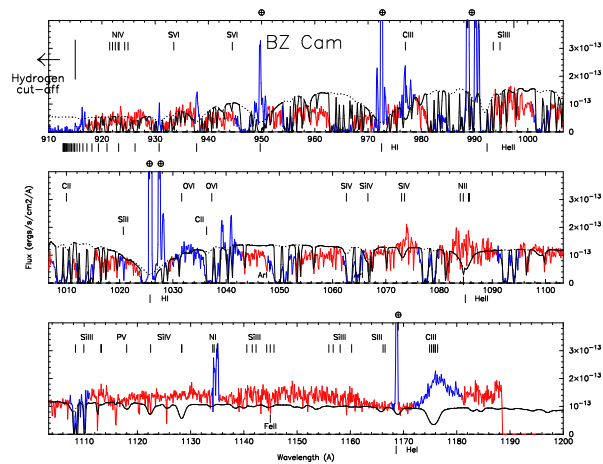}                       
\caption{
The {\it FUSE} spectrum of BZ Cam (in red) is fitted
with a hot WD to agree with the distance to the system as 
well as with the flux in the short wavelengths. 
The WD model (including the ISM absorption model) 
is drawn as the solid black line,
it has $M_{\rm wd}=0.4M_{\odot}$,  
$T_{\rm wd}=45,000$K, solar composition and $V_{rot} \sin{(i)} =200$km/s 
(the same WD model without the ISM absorption model is shown with 
the dotted black line). 
The distance obtained
from the fit is 791pc.  The ISM absorption features as well as airglow
emission lines are marked in blue.  We have marked
some of the prominent absoprtion lines that are usually observed in 
accreting WD with a high temperature, these lines are not observed
here, a sign that the observed spectrum might have some emission lines 
on top of the absorption lines or that the photospher has a low
metalicity. 
} 
\end{figure}

\paragraph{Modeling the {\it IUE} Spectrum}

The {\it IUE} spectrum of BZ Cam was modeled with a 12,500K Kurucz model 
with $\log{g}=2.5$ by \citet{pri00}, indicating that  
a standard disk model does not fit the spectrum.
Therefore, we do not attempt a standard disk model but instead we 
use a modified disk model. 

We assume a WD mass of $0.4M_{\odot}$, a low inclination ($i=10^{\circ}$ 
\& $i=20^{\circ}$) and we vary the mass accretion rate from 
$10^{-10}M_{\odot}$/yr to $10^{-8}M_{\odot}$/yr. We assume 
a WD radius $R_{\rm wd}=1.15 \times 10^9$cm, actually 
corresponding a 10,000K $0.4M_{\odot}$ WD (see section 3.2), and we vary the 
inner disk radius $R_0$ from $1 R_{\rm wd}$ to $3 R_{\rm wd}$. 
We keep the outer part of the
disk at an isothermal temperature of 10, 11, or 12,000K. 
 
We find the best modified disk model fit  
has a mass accretion rate $\dot{M}=10^{-8.5}M_{\odot}$/yr, $i=20^{\circ}$, 
a disk inner radius $R_0=2R_{\rm wd}$ (=23,000km),                    
and the outer half of the disk is set to an insothermal temperature 
$T=10,000$K. This outer half of the disk has a surface area 3 times
larger than the inner half of the disk, making this modified disk
model resemble an isothermal disk model.
However, in order for this modified disk 
model to agree with the known distance ($830\pm160$pc),       
the outer disk radius had to be 
set at $\sim 0.5 \pm 0.1 a$, corresponding to $R_d \approx 22R_{\rm wd}$. 
Though such a disk radius is about twice the accepted radius (0.3a)
in a CV binary,
it is still within the limits of the Roche lobe. 
Chosing an outer disk radius smaller than this value, decreases the scaled 
distances.  The fit is shown in Fig.11. 

Since this modified disk model is close to an isothermal disk, we 
try next an isothermal disk model, with the same inner and outer 
radii. The isothermal disk model 
gives very similar results when its temperature is set to 
11,000K. 
In this model, shown in Fig.12, the defficiency in flux in the
shorter wavelength range (the Ly$\alpha$ region) is a little more pronounced
than in the modified disk model.
In the isothermal disk model, the exact value of the inner disk radius, 
i.e. whether it is 1 or 2 $R_{\rm wd}$, has very little effect on the 
results as  it only changes the surface area of the disk by a tiny fraction.   
Also, the value of the WD mass is not taken into account.  
However, when integrating the luminosity over the surface of the isothermal  
disk, the resulting mass accretion rate corresponds to  
$3.7 \times 10^{-9}M_{\odot}$/yr for a $0.4M_{\odot}$ mass WD  
and  
$2.7 \times 10^{-9}M_{\odot}$/yr for a $0.7M_{\odot}$ mass WD.

\begin{figure}[h] 
\vspace{-9.cm} 
\plotone{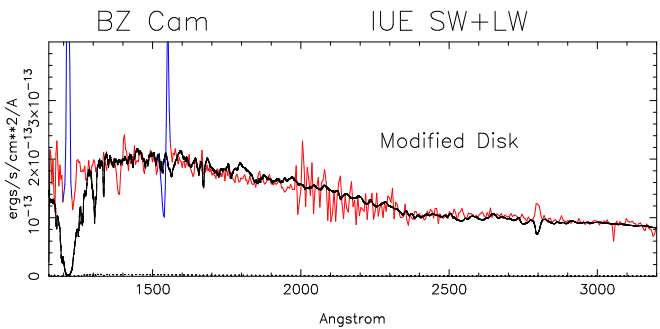}                       
\caption{
The {\it IUE} spectrum of BZ Cam (in red, with masked emission
lines in blue), is modeled with a modified disk model (in black). The WD mass
is $0.4M_{\odot}$, the disk has a mass accretion rate of 
$\dot{M}=10^{-8.5}M_{\odot}$/yr, an inclination $i=20^{\circ}$, and the
outer half of the disk has been set to an isothermal temperature
$T=10,000$K. The inner disk radius is $R_0=2R_{\rm wd}$ (zero temperature
WD), and the outer disk radius is set to $0.5$ the binary separation 
(within the limits of the Roche lobe) 
in oder to match the distance to the system. 
Note that within the 
Ly$\alpha$ region the model seems to be defficient in flux.  
} 
\end{figure}

\begin{figure}[h] 
\vspace{-9.cm} 
\plotone{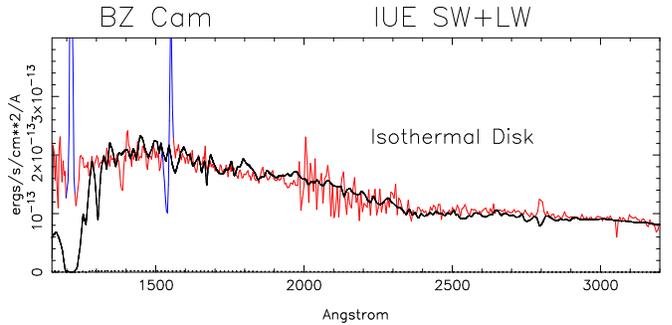}                       
\caption{
Same as in the previous figure, but now the spectrum is modeled
with an isothermal disk only. The disk has a temperature of 
11,000K. Here too, in order to scale the model to the known distance, 
the radius of the disk has to be $0.5a$. 
departing the most from a standard disk model.  The defficiency 
in flux in the 
Ly$\alpha$ region is even more pronounced than in the modified 
disk model. 
} 
\end{figure}

{\bf Self-Consistent Modeling of the {\it FUSE} \& {\it IUE} Spectra} 

Since the {\it FUSE} and {\it IUE} spectra were obtained years apart,
in slightly different brightness states, 
and with different telescopes/instruments, 
we do not expect them to match and we do not attemp to combine them. 
However, when modeling each spectrum separately ({\it IUE} or {\it FUSE}),  
we must take into consideration the results obtained 
from the modeling of the other spectrum ({\it FUSE} or{\it IUE} respectively).  
Namely, we now model the {\it IUE} spectrum with a cold disk 
{\it and} a hot
WD model, similar but not identical, to the WD model obtained from the modeling 
of the {\it FUSE} spectrum;  
and we model the {\it FUSE} spectrum with a hot WD {\it and} 
a cold disk 
model, similar but not identical, to the cold disk model obtained from the modeling 
of the {\it IUE} spectrum. 

For the disk model, 
we choose the isothermal disk model as it is easier to 
parametrize, and gives results similar to the modified disk model. 
Namely, the WD mass does not enter the disk model 
and the WD radius, taken as the inner radius of the disk, has minimum
effect on the results. 
For the WD model, we choose a $0.4M_{\odot}$ WD 
with a radius increasing with temperature. 
As we have mentioned earlier, 
only the WD radius and temperature affects the results.

For the {\it FUSE} spectrum we find that the 45,000K WD model,  
with a radius of 18,540km, 
is improved significantly with  
the addition of a cold isothermal disk with T=11,000K. 
The cold disk contributes mainly in the longer wavelengths of
the {\it FUSE} spectrum and helps improve the fit in that region 
(see Fig.13). 
The disk has an outer radius of 0.5 the binary 
separation and the distance obtained from the fit is 854pc.   
The mass accretion rate obtained by integrating the luminosity
of the disk, is $3.7 \times 10^{-9} M_{\odot}$/yr, and the heated
WD luminosity is of the same order.  

The fit is improved further by increasing the WD temperature 
to 50,0000K and the isothermal disk temperature to 12,000K, however the 
distance increases to 1137pc. This model is shown in Fig.14.   
The disk integrated luminosity gives a mass accretion 
$\dot{M}=5 \times 10^{-9}M_{\odot}$/yr and here too the WD luminosity is
of the same order.    

For the {\it IUE} spectrum we find that for the model to fit the spectrum, 
the temperature of the disk has to be decreased as the temperature of
the WD increases. The best fit (Fig.15) is obtained for a disk with a temperature  
of 10,000K combined with a 40,000K WD. This model also gives a
distance of the order of 650pc for a an outer disk radius $\sim 0.6a$. 
The integrated luminosity of the disk gives a mass accretion rate of 
$2.5 \times 10^{-9}M_{\odot}$/yr, while the WD luminosity is about 10 
percent less than that of the disk.  

The results of the analysis of the UV spectra of BZ Cam can be summarized
as follows: a $0.4M_{\odot}$ WD with a temperature of 45,000$\pm$5,000K, 
and a disk with a temperature of 11,000$\pm$1000K, extending to 
$0.5 \pm 0.1$ the binary separation. Taking only the disk luminosity
into consideration gives a mass accretion rate of the order of 
$2.5 \times 10^{-9}M_{\odot}$ to $5 \times 10^{-9}M_{\odot}$/yr, 
and twice as large when also taking the WD luminosity contribution
(assuming that the WD is heated due to the advection of energy 
from the inner disk). The WD contributes mainly to the {\it FUSE} spectral range
while the cold disk contributes mainly to the {\it IUE} spectral range. 
The WD mass could be larger, i.e. $0.7M_{\odot}$,
but its radius would have to be of the order of 20,000km, i.e. twice as large
as expected for such a mass (again, see section 3.2 for details).

\begin{figure}[h] 
\vspace{-2.cm} 
\plotone{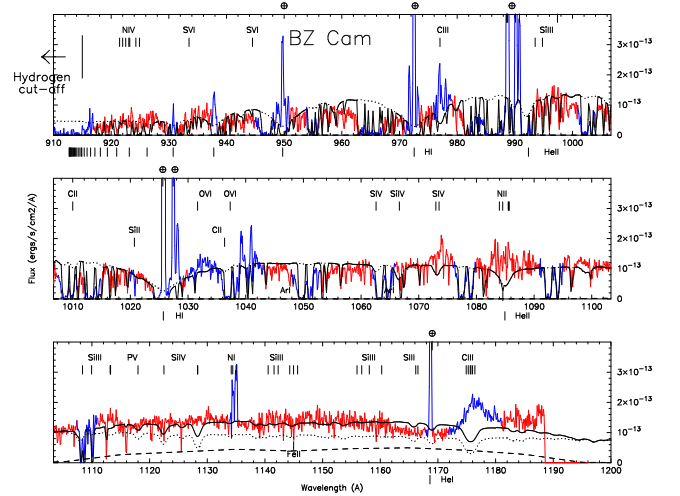}                       
\caption{
The {\it FUSE} spectrum of BZ Cam is modeled with $0.4M_{\odot}$ 45,000K WD 
(dotted line) plus a 11,000K isothermal disk 
(dashed line) of radius 0.5a. 
The combined model (solid black line) includes a basic ISM absorption
model.
  Note that the isothermal
disk, contributing only 7\% of the    flux in the {\it FUSE} range, 
improves the overall fit and more particularly in the 
longer wavelength. The distance obtained from the fit is 854pc. 
The region of the C\,{\sc iii} (1160-1180) shows a possible 
P-Cygni profile, though it is also a region exhibiting a  
detector artifact  known as the `worm'.  
} 
\end{figure}

\begin{figure}[h] 
\vspace{-2.cm} 
\plotone{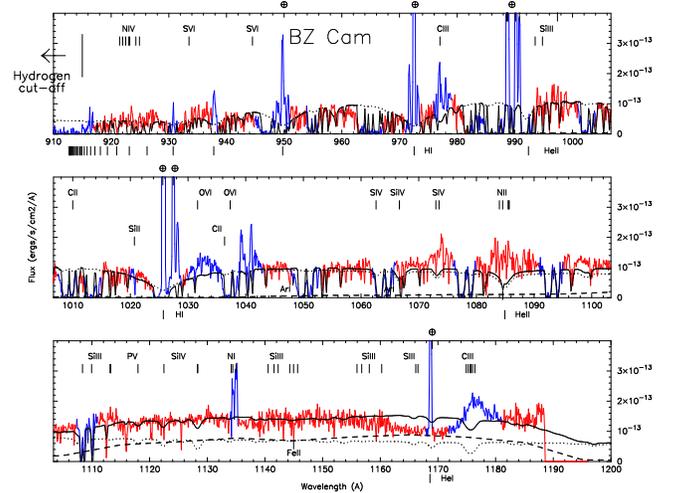}                       
\caption{
The {\it FUSE} spectrum of BZ Cam is modeled with $0.4M_{\odot}$ 50,000K WD 
(dotted line) plus a 12,000K isothermal disk (dashed line). 
The combined model (solid black line) includes a basic ISM absorption
model.   
The isothermal
disk now contributes 25\% in the {\it FUSE} range
of the flux, and further improves the fit in the 
longer wavelength. The distance obtained from the fit is 1137pc. 
} 
\end{figure}

\begin{figure}[h] 
\vspace{-4.cm} 
\plotone{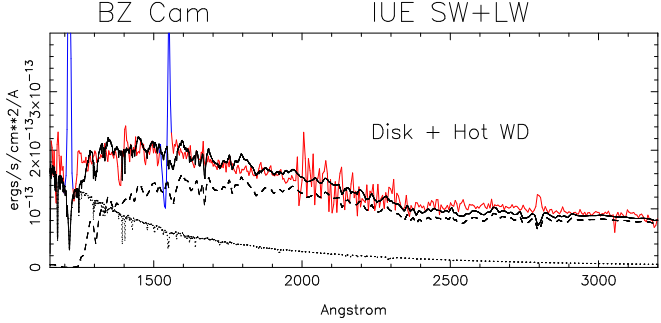}                       
\vspace{-1.cm} 
\caption{
The {\it IUE} spectrum of BZ Cam is modeled 
(solid black line) with
an isothermal disk plus a hot WD. The disk (dashed black line) 
has a temperature of 10,000K
and the WD (dotted black line) has a temperature of 40,000K.  
Note that the WD contribution greatly improves the fit in the Ly$\alpha$
region. 
} 
\end{figure}

\subsection{V592 Cas} 

We assumed a $0.8M_{\odot}$ WD mass with a radius $R_{\rm wd}= 7,000$km, 
which is consistent with a temperature of such a WD at about 16,000K.  
The inner radius of the disk, $R_0$, was placed at $1.0 R_{\rm wd}$,    
$1.2 R_{\rm wd}$,    $1.5 R_{\rm wd}$, and     
$2.0 R_{\rm wd}$. 
The outer radius was placed at $36R_{\rm wd} \approx a/3$, where
$a$ is the binary separation of V592 Cas derived from Table 1. 
The outer region of the disk was allowed to 
take different temperatures: either the actual disk 
temperature, or 10,000 K, 11,000 K, or 12,000 K.  
The size of that outer region was also varied. 
Some preliminary modelings 
were carried out using a grid of standard disk models, to obtain
an order of estimate of the mass accretion rate, which gave
 $\dot{M} \sim 10^{-8} M_{\odot}$/yr. Modified disk models were then
computed within the vicinity of this mass accretion rate assuming 
an inclination angle of $30^{\circ}$.  

The models were chosen to provide the best fit and the
correct distance ($\approx 350$ pc) to the source. 
The best fit is for a modified disk with a mass accretion rate of 
$\dot{M}=10^{-8.2}M_{\odot}$/yr ($\sim 6.3 \times 10^{-9} M_{\odot}$/yr), 
an inner disk radius placed at $R_0=1.2R_{\rm wd}$, 
an outer disk radius at 30$R_0$, and an isothermal outer disk
between $r=20R_0$ and $r=30R_0$ with a temperature T=12,000K. 
A 45,000K WD contribution was added to the model for completeness
\citep{hoa09} 
but contributed only a few percent to the flux and did not 
affect the model. Consequently, the temperature of the WD could
not be assessed from our modeling. The fit is presented in Fig.16
in the {\it FUSE} spectral range and in Fig.17 in the {\it IUE} SWP+LWP 
spectral range.   

For a 16,000K WD with a radius of 7,000km, the modified disk model, 
with a 8,400km inner radius,  
starts only at $1.2R_{\rm wd}$, and for a 45,000K WD with a radius 
of 7,450km the disk starts at $1.13R_{\rm wd}$. In either case 
the size of the BL would be  relatively narrow ($\sim 0.1-0.2R_{\rm wd}$)
and this modified disk model is very similar to a standard disk 
model in its inner region, and differs mainly in the assumption of an 
outer isothermal region. V592 Cas is not a case as extreme as 
BZ Cam or MV Lyr.

\begin{figure}[h] 
%\vspace{-6.cm} 
\plotone{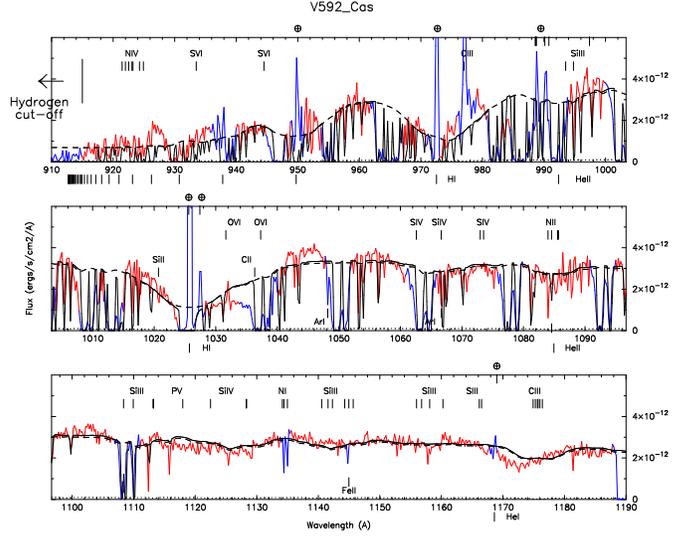}                       
\caption{
The {\it FUSE} spectrum of V592 Cas (in red) has been fitted with a combined WD + 
modified disk model (solid black line). The WD model, dotted line,
has a mass of $0.8M_{\odot}$ with a temperature of 45,000K, but
contributes very little to the combined model and was added only
for completeness. The modified disk model, dashed line, has a mass accretion 
rate of $6.3 \times 10^{-9} M_{\odot}$/yr ($\log(\dot{M})=-8.2$), with an
inclination of $30^{\circ}$. The inner radius of the disk is set 
at $R_0 = 1.2 R_{\rm wd}$ (8,400 km), and the outer radius is 
set at 30$R_0$.
The outer region of the disk between $20R_0$ and $30R_0$ was  
set to a temperature of 12,000 K. This model gives a distance of 342 pc.  
The regions affected by ISM absorption features and sharp
emission lines (due to day light) are omitted (masked) for
the fitting and have been colored in blue. 
The location of absorption lines observed in the {\it FUSE} spectra of some
CVs have been marked, but are not especially observed here.  
} 
\end{figure}

\begin{figure}[h] 
\vspace{-9.cm} 
\plotone{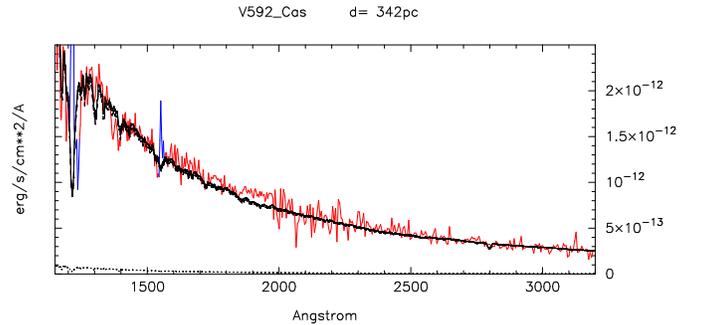}                       
\caption{The same disk model shown in Fig.16 is now fitted to the
{\it IUE} spectrum of V592 Cas. 
For clarity the vertical scale is different than in
Fig.16.  
} 
\end{figure}

\vspace{1.cm}          
\section{{\bf Discussion}} 

In the present UV spectral analysis, we find that the data for the UX UMa system
V592 Cas are consistent  
with a slightly modified disk model (very similar to 
a standard disk) with an inner disk radius at $R_0=1.2R_{\rm wd}$, 
$\dot{M} = 10^{-8.2}M_{\odot}$/yr,  
and an outer 12,000K isothermal disk. The small surface of the 
$\sim 0.8M_{\odot}$ 
WD is overshone by the disk, as its contribution, 
even at a temperature $T_{\rm wd}=45,000$K, is negligible. 
The data for our two VY Scl systems MV Lyr \& BZ Cam point to a 
different scenario. 
For MV Lyr we find a significantly modified disk model
with an inner radius as large as $2R_{\rm wd}$, 
and an outer 10,000K isothermal disk. The contribution of the
45,000K $0.8M_{\odot}$ WD is significant.
BZ Cam is an even more extreme case in that it has a slope shallower
than MV Lyr and V592 Cas, and its spectrum is consistent with 
a large cold isothermal disk 
with $T\sim 10-11,000$K extending to the limit of the Roche
lobe ($\sim 0.5a$).  A modified disk model with an inner radius
of $2R_{\rm wd}$ (40,000km) and an extended outer isothermal
region provides the same solution. 
Its $0.4M_{\odot}$ WD with  a 20,000km radius, and a temperature of 
45,000K contributes most of the flux in the {\it FUSE} range. 
For both MV Lyr and BZ Cam the mass accretion rate is about twice
as small as for V592 Cas. 

The hot WD scenario in the two VY Scl NL systems MV Lyr and BZ Cam,  
with an inner disk hole $\sim 2R_{\rm wd}$,  
is consistent with our recent {\it Swift} X-ray data 
analysis which reveals that the optically thin hard X-ray 
emitting boundary layer is under-luminous, a sign that it 
cannot radiate its energy to cool efficiently, 
and, instead, the boundary layer energy is advected into the outer layer of 
the white dwarf, heating it, possibly inflating its radius and
launching an outflow.  
This is the self-consistent global picture we draw from our present UV   
spectral analysis and our previous {\it Swift} X-ray 
analysis. In this picture, the optically thin BL extends to 
$r \sim 2R_{\rm wd}$ and heats the WD through advection of energy. 
For V592 Cas, the inner hole is much smaller extending only to 
$\sim 1.2R_{\rm wd}$.  

To further check this scenario, we select a larger and self-consistent 
set  of X-ray data for CVs to compare to the archival {\it IUE} 
data. If the above scenario is correct, we expect the systems with
UV spectra departing
from the standard disk model to have a harder X-ray emission 
than those systems with UV spectra agreeing with the standard disk model.   
For this purpose, we turn to \citet{van96}, 
who  computed the X-ray hardness ratio from ROSAT observations
of non-magnetic cataclysmic variables, including 5 VY Scl NLs and 
6 UX UMa NLs observed in a state of high accretion rate, 
and a number of DN systems observed in their quiescent state. 
ROSAT has an energy range (0.1-2.4keV) that is {\it relatively} soft 
considering that some intermediate polars can be as hard as 20-300keV.  
In spite of the fact that the DNe were observed in quiescence, 
\citet{van96} found that the NL VY systems 
had a X-ray hardness ratio larger than the NL UX UMa
and quiescent DN SU UMa systems. 
DN systems in outburst are known to exhibit hard X-rays too
(only a few systems exhibit EUV/soft X-rays). 
During outburst, however, the X-ray temperature is lower than in quiescence
\citep{bal15}. Consequently, the hardness ratio of the quiescent DNe computed 
in \citet{van96} is an upper limit to hardness ratio of these same
DNe in outburst. So the data in \citet{van96} give us the hardness
ratio of disk-dominated UX and VY NLs and an upper limit to the hardness ratio 
of DN in outburst. 
 
As noted by 
\citet{lad91}, disk-dominated NL systems  
have a shallower UV continuum slope than the DN in outburst
and cannot be modeled with standard disk models. 
To {\it quantitatively} assess the departure from the standard disk model
in the UV, we measure the slope of the continuum of the {\it IUE} spectra,  
flux in $<$erg/s/cm$^2$/\AA$>$ as a function of wavelength in $<$\AA$>$
on a log-log scale, for the sample of non-magnetic disk systems 
in \citet{van96} (for which {\it IUE} data exist). Using the LWP segment
and the longer wavelength region of the SWP spectrum 
past the 1550\AA\ carbon line, we obtain a UV spectral coverage 
from $\sim 1600$\AA\ to 3,200\AA . We deredden all the spectra 
based on the E(B-V) values from the litterature \citep{bru94} or
directly from the 2175\AA\ absorption feature using the extinction
curve from \citet{fit07} with $R=3.1$. 
We then measure the slope of the UV continuum
directly from the log-log graphs that we generate for these systems
(graphs similar in appearance to Figs. 1, 2, \& 3). 
We have here a limited sample of systems as we decided
to take only those systems for which the hardness ratio was measured from
the same telescope in the same manner and \citet{van96} provides
such a self-consistent sample.  

In Table 3, we list the ROSAT 
X-ray hardness ratio and {\it IUE} UV continuum slope for the  
9 DN systems, 6 UX NL systems and 4 VY NL systems together with the
novalike V795 Her and the two old novae V603 Aql, and RR Pic. 
V592 Cas was not included in \citet{van96} and has no hardness
ratio listed. 
MV Lyr is listed twice, once for its intermediate 
state and once for its high state that exhibits a slightly different UV 
continuum slopes.  
At the bottom of the table we also list the UV continuum slope of 
some standard disk models computed with TLUSTY and SYNSPEC 
(e.g. \citet{wad98}) for comparison. The standard disk models have a slope
that flattens as the mass accretion rate decreases.

\begin{deluxetable}{lccc}
\tablewidth{0pt}
\tablecaption{System Parameters}
\tablehead{
System    &  Type &  UV     &    hardness  \\     
 Name     &        & slope  &    ratio      
}
\startdata
 WZ Sge    &  DN/SU   & -2.4    &    0.21  \\   
 SW UMa    &  DN/SU   & -2.5    &    0.00  \\   
 SU UMa    &  DN/SU   & -1.9    &    0.24  \\   
 Z Cha     &  DN/SU   & -1.7    &    0.53  \\   
 WX Hyi    &  DN/SU   & -2.0    &    0.46  \\    
 T Leo     &  DN/SU   & -2.3    &    0.22  \\  
 VW Hyi    &  DN/SU   & -2.4    &    0.05  \\ 
 AB Dra    &  DN/UG   & -2.1    &    0.84  \\  
 WW Cet    & DN/ZC  &  -1.75  & 0.50       \\ 
 RW Tri    &  NL/UX   & -1.8    &    1.39  \\ 
 AC Cnc    &  NL/UX   & -1.7    &   -0.26  \\ 
 UX UMa    &  NL/UX   & -2.0    &    0.05  \\ 
 IX Vel    &  NL/UX   & -2.2    &    0.10  \\ 
 V3885 Sgr &  NL/UX   & -2.1    &    0.38  \\ 
 V592 Cas  &  NL/UX   & -2.4    &    ---   \\ 
 BZ Cam    &  NL/VY   & -1.15   &    0.95  \\ 
 KR Aur    &  NL/VY   & -1.55   &    0.96  \\ 
 MV Lyr i  &  NL/VY   & -1.4    &    0.83  \\ 
 MV Lyr h  &  NL/VY   & -1.6    &    0.83  \\ 
 TT Ari    &  NL/VY   & -1.9    &    0.83  \\ 
 V795 Her  & NL/SH  &  -1.9   & 0.00     \\ 
 V603 Aql  & CN/SH  &  -2.35  & 0.83        \\   
 RR Pic    & CN/SW  &  -1.95  & -0.49       \\ 
\hline 
System    & $\dot{M}$         &  UV     &    hardness  \\     
          & $<M_{\odot}/yr>$  &  slope  &    ratio     \\  
\hline 
Standard Disk   & $10^{-9.5}$ &  -2.52   &    ---   \\ 
Standard Disk   & $10^{-9.0}$ &  -2.89   &    ---   \\ 
Standard Disk   & $10^{-8.5}$ &  -2.88   &    ---   \\ 
Standard Disk   & $10^{-8.0}$ &  -2.94   &    ---   \\ 
\enddata  
\tablecomments{The subtypes are as in \citet{rit03}, except for the novae
that are denoted CN (classical nova). 
The UV continuum slope was measured directly from the dereddened {\it IUE}
spectra 
and the ROSAT X-ray hardness ratio was taken from  \citet{van96}.
Some standard disk models \citep{wad98} are also included for 
comparison.} 
\end{deluxetable}

In Fig.18 we plot the UV continuum slope of these systems 
against the ROSAT X-ray hardness ratio reported in \citet{van96}.   
Eventhough the X-ray hardness ratio for DNe is an upper limit, and all the
DNe (green triangle) are expected to be located more to the left,  
had the hardness ratio been measured in outburst, 
there is a clear separation between DNe and VY Scl systems. 
Furthermore, for this limited sample, it appears that 
the VY Scl systems have a shallower UV slope with harder X-ray spectrum,  
consistent with the scenario of an optically thin inner disk 
with a modified disk model. The DNe have a softer spectrum
and a steeper UV slope consistent with either a standard disk model
or a less modified disk model
with a less extended optically thin inner disk (e.g. as we found 
for V592 Cas).  
There is an overall tendancy pointing to the
fact that the more disk-dominated CVs depart   
from the standard disk model in the UV (shallower slope), the  
harder is their X-ray emission, respectively starting with the DN systems 
and ending with the VY Scl systems.

\begin{figure}[h] 
\vspace{-1.5cm} 
\plotone{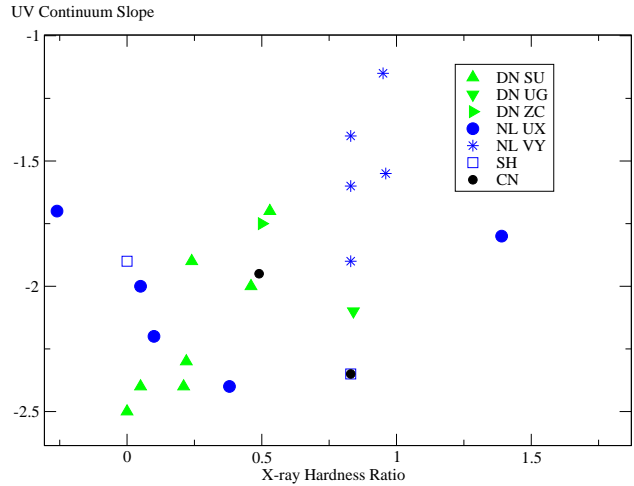}                       
\caption{
The UV continuum slope of DNe in outburst and NLs in high state
(disk-dominated CVs) 
is plotted against the measured X-ray hardness ratio 
of quiescent DNe and NLs in high state for the systems listed
in Table 3.
The subtypes are as indicated in the upper right panel. 
V603 Aql is here taken both as a CN and as a NL SH. 
V592 Cas was not included as it is not listed in \citet{van96}. 
DNe in outburst have softer X-ray than in quiescence 
\citep{bal15}, implying
that the DNe (green triangles) would shift to the left if the hardness
ratio had been measured during outburst. Even so, there is a clear
separation between DNe and VY NLs. 
On the overall, for disk-dominated CVs, the harder is the X-ray,  
the shallower is the slope of of the UV continuum. 
The departure from the standard disk model in 
the UV correlates with the X-ray hardness 
of the system.
} 
\end{figure}

Motivated by this findings, and to further check the nature of VY Scl systems,
we decided to measure the
UV continuum slope for all the available {\it IUE} archive of 
disk-dominated CVs.  We found 105 CVs with {\it IUE} spectra
taken when the disk-dominated the system 
and for which a continuum slope could be measured. 
Reddening  values E(B-V) were taken 
from \citet{bru94} when available, or were directly 
assessed using the 2175\AA\ feature. 
The UV continuum slope for all the systems was measured after the spectra 
were dereddened. 
For each system in Table 4, 
we list the resulting UV slope  
together with the Data ID, reddening  and orbital period.  
Some systems with only SWP spectra available were included 
(V794 Aql and CZ Ori), others were not included 
because the value of the slope could not be assessed (e.g. V380 Oph). 
%or the slope in that spectral was zero (e.g. UU Aqr). 
Some systems did not reveal any obvious continuum
from which a slope value could be derived,
as these spectra were affected by noise, emission lines, 
or possible additional components (e.g. BH Lyn, V348 Pup, 
DW UMa).
For GI Mon, the only two {\it IUE} spectra (SWP \& LWP) 
gave a different
slope: SWP gave -1.5 and LWP gave -1.1, and the value listed is their
average. 

We plot the UV continuum slope against the orbital 
period for all the CVs in Table 4 in two Figures:  
we show in Fig.19 the systems above the period gap,   
and in Fig.20 we show the systems below the period gap
(for clarity the $X$-axis has been stretched in Fig.20).

\begin{small} 
\begin{deluxetable*}{lcccllcc}
\tablewidth{0pt}
\tablecaption{UV Continuum Slope of Disk-Dominated Cataclysmic Variables}
\tablehead{
System    & Type &   IUE    &   IUE   &  Period   &  UV   & E(B-V)  & $i$ \\
Name      & Subtype & Short  &   Long  &  $<d>$    &  slope &       &$<$deg$>$ 
}
\startdata
WZ Sge    & DN/SU  & SP03507  & LR03086 & 0.056694  & -2.4   & 0.00 & 77  \\  
SW UMa    & DN/SU  & SP27871  & LP07754 & 0.056815  & -2.5   & 0.00 & 45  \\  
WX Cet    & DN/SU  & SP36511  & LP15730 & 0.05827   & -2.3   & 0.00 & --- \\  
CC Scl    & NL/IP  & SP34427  & LP14189 & 0.0584    & -2.0   & 0.00 & 81   \\ 
T Leo     & DN/SU  & SP33646  & LP13312 & 0.05882   & -2.3   & 0.00 & 65    \\  
CP Pup    & CN/SH  & SP27806  & LP07782 & 0.061264  & -2.0   & 0.20 & ---   \\ 
V1159 Ori & DN/SU  & SP56781  & LP31957 & 0.06218   & -2.2   & 0.00 & --- \\  
V436 Cen  & DN/SU  & SP54246  & LP30319 & 0.062501  & -2.7   & 0.07 & 65  \\  
BC UMa    & DN/SU  & SP50665  & LP28027 & 0.06261   & -2.2   & 0.00 & --- \\  
EK TrA    & DN/SU  & SP09705  & LR08446 & 0.06288   & -2.25  & 0.03 & 58  \\  
TV Crv    & DN/SU  & SP41842  & LP20597 & 0.0629    & -2.6   & 0.10 & --- \\  
OY Car    & DN/SU  & SP25857  & LP05906 & 0.063121  & -2.2   & 0.05 & 83  \\ 
VY Aqr    & DN/SU  & SP21719  & LP02366 & 0.06309   & -2.2   & 0.00 & --- \\  
ER UMa    & DN/SU  & SP40947  & LP19846 & 0.06366   & -2.05  & 0.00 & --- \\  
EX Hya    & NL/IP  & SP03858  & LR03435 & 0.0682338 & -1.6   & 0.00 & 78   \\ 
IR Gem    & DN/SU  & SP38524  & LP17696 & 0.0684    & -2.15  & 0.00 & --- \\ 
VZ Pyx    & DN/SU  & SP44128  & LP22531 & 0.07332   & -2.35  & 0.07 & --- \\ 
AY Lyr    & DN/SU  & SP09342  & LR08100 & 0.0737    & -2.4   & 0.00 & --- \\ 
VW Hyi    & DN/SU  & SP48680  & LP26405 & 0.074271  & -2.4   & 0.00 & --- \\ 
Z Cha     & DN/SU  & SP30677  & LP10466 & 0.074499  & -1.7   & 0.00 & 82  \\ 
WX Hyi    & DN/SU  & SP23952  & LP04221 & 0.074813  & -2.0   & 0.01 & 40  \\ 
T Pyx     & CN     & SP33034  & LP12791 & 0.076223  & -2.6   & 0.35 & --- \\ 
SU UMa    & DN/SU  & SP34824  & LP14532 & 0.07635   & -1.9   & 0.00 & --- \\ 
YZ Cnc    & DN/SU  & SP03727  & LR03308 & 0.0868    & -1.85  & 0.00 & 38  \\ 
V795 Her  & NL/SH  & SP22901  & LP03269 & 0.1082648 & -1.9   & 0.00 & ---  \\ 
V592 Cas  & NL/UX  & Table 2  & Table 2 & 0.115063  & -2.4   & 0.22 & 28  \\ 
TU Men    & DN/SU  & SP10665  & LR09374 & 0.1172    & -2.6   & 0.08 & --- \\ 
V442 Oph  & NL/VY  & SP14731  & LR11298 & 0.12433   & -2.2   & 0.20 & --- \\ 
LQ Peg    & NL/VY  & SP17367  & LR13618 & 0.124747  & -2.2   & 0.00 & --- \\  
AH Men    & NL/SW  & SP43037  & LP21666 & 0.12721   & -1.85  & 0.12 & --- \\  
DN Gem    & CN     & SP38213  & LP17402 & 0.127844  & -1.7   & 0.10 & --- \\ 
KQ Mon    & NL/UX  & SP15384  & LR11915 & 0.128     & -1.8   & 0.04 & --- \\ 
MV Lyr    & NL/VY  & SP07296  & LR06288 & 0.132335  & -1.5   & 0.00 & 12  \\ 
SW Sex    & NL/SW  & SP21533  & LP02262 & 0.134938  & -1.0   & 0.00 & $>$75 \\ 
HL Aqr    & NL/SW  & SP23325  & LP03647 & 0.13557   & -2.15  & 0.05 & --- \\ 
TT Ari    & NL/VY  & SP42147  & LP20922 & 0.137550  & -1.9   & 0.03 & ---    \\ 
V603 Aql  & CN/SH  & SP05758  & LR04994 & 0.138201  & -2.35  & 0.08 & 13   \\   
WX Ari    & NL/SW  & SP55953  & LP31494 & 0.139351  & -1.0   & 0.00 & 72     \\  
V1315 Aql & NL/SW  & SP27096  & LP07085 & 0.13969   & -1.1   & 0.10 & 82    \\ 
V1223 Sgr & NL/IP  & SP13365  & LR10022 & 0.140244  & -2.15  & 0.15 & 24   \\ 
V2400 Oph & NL/IP  & SP47007  & LP24971 & 0.142     & -2.4   & 0.40 & ---  \\ 
LN UMa    & NL/SW  & SP40948  & LP19847 & 0.1444    & -1.9   & 0.15 & --- \\ 
V751 Cyg  & NL/VY  & SP25774  & LP05819 & 0.144464  & -1.1   & 0.20 & --- \\ 
RR Pic    & CN/SW  & SP06625  & LR05687 & 0.145025  & -1.95  & 0.00 & 65   \\ 
PX And    & NL/SW  & SP39273  & LP18414 & 0.146353  & -1.    & 0.05 & --- \\ 
V533 Her  & CN/IP? & SP44805  & LP23205 & 0.147     & -1.4   & 0.03 & ---  \\  
V425 Cas  & NL/VY  & SP15267  & LR11783 & 0.1496    & -1.5   & 0.10 & 25  \\ 
AO Psc    & NL/IP  & SP09706  & LR08447 & 0.1496252 & -2.2   & 0.10 & ---  \\
AB Dra    & DN/UG  & SP17619  & LR13886 & 0.1520    & -2.1   & 0.10 & --- \\ 
V794 Aql  & NL/VY  & SP50754  &  ---    & 0.1533    & -2.1   & 0.10 & --- \\ 
BP Lyn    & NL/SW  & SP32940  & LP12690 & 0.152812  & -2.1   & 0.00 & ---   \\ 
BZ Cam    & NL/VY  & Table 2  & Table 2 & 0.15353   & -1.15  & 0.05 & --- \\ 
LX Ser    & VY/SW  & SP08070  & LR07037 & 0.158432  & -1.0   & 0.00 & 90  \\ 
CY Lyr    & DN/UG  & SP21030  & LR16779 & 0.1591    & -2.5   & 0.15 & --- \\ 
CM Del    & NL/UX  & SP14707  & LR11282 & 0.162     & -1.9   & 0.09 & 73  \\ 
%\enddata  
%\end{deluxetable*}
%\end{small} 
%
%
%
%\begin{small} 
%\begin{deluxetable*}{lcccllcc}
%\tablenum{4} 
%\tablewidth{0pt}
%\tablecaption{Continue}
%\tablehead{
%System    & Type &   IUE    &   IUE   &  Period   &  UV   & E(B-V)  & $i$ \\
%Name      & Subtype & Short  &   Long  &  $<d>$    &  slope &       &$<$deg$>$ 
%}
%\startdata
KT Per    & DN/UG  & SP17712  & LR13968 & 0.16265777 &-2.7   & 0.20 & --- \\  
KR Aur    & NL/VY  & SP14734  & LR11299 & 0.16280   & -1.55  & 0.05 & 38  \\ 
AR And    & DN/UG  & SP18877  & LR14885 & 0.16302   & -2.4   & 0.02 & --- \\ 
CN Ori    & DN/UG  & SP32593  & LP12364 & 0.163199  & -2.15  & 0.00 & 67  \\ 
X Leo     & DN/UG  & SP15951  & LR12282 & 0.1646    & -2.4   & 0.00 & --- \\ 
VW Vul    & DN/ZC  & SP18875  & LR14883 & 0.16870   & -2.1   & 0.15 & --- \\ 
UZ Ser    & DN/ZC  & SP15078  & LR11605 & 0.173     & -3.0   & 0.25 & --- \\  
V405 Aur  & NL/IP  & SP56765  & LP31949 & 0.17345   & -0.5   & 0.15 & $<$5   \\ 
LS Peg    & NL/IP  & SP39782  & LP18948 & 0.174774  & -1.7   & 0.04 & 30   \\ 
WW Cet    & DN/ZC  & SP10664  & LR09373 & 0.1758    & -1.75  & 0.03 & 54     \\ 
U Gem     & DN/UG  & SP10327  & LR08987 & 0.17690619 &-2.5   & 0.04 & 70     \\ 
GI Mon    & CN/IP  & SP38419  & LP17709 & 0.1802    & -1.3   & 0.10 & ---  \\ 
SS Aur    & DN/UG  & SP18610  & LR14679 & 0.1828    & -2.4   & 0.08 & 38  \\ 
DQ Her    & NL/IP  & SP09201  & LR07981 & 0.19362   & -0.75  & 0.08 & 90   \\ 
IX Vel    & NL/UX  & SP18578  & LR14657 & 0.193927  & -2.2   & 0.01 & 57  \\ 
UX UMa    & NL/UX  & SP46816  & LP24773 & 0.196671  & -2.0   & 0.02 & 70  \\ 
FO Aqr    & CN/IP  & SP15260  & LR11774 & 0.202     & -0.5   & 0.00 & ---  \\ 
V825 Her  & NL/UX? & SP17353  & LR13601 & 0.206     & -2.0   & 0.00 & ---     \\   
HX Peg    & DN/ZC  & SP37458  & LP16654 & 0.2008    & -2.0   & 0.00 & --- \\ 
V3885 Sgr & DN/UX  & SP25655  & LP05709 & 0.207161  & -2.1   & 0.02 & --- \\ 
RX And    & DN/ZC  & SP17673  & LR13942 & 0.209893  & -2.1   & 0.02 & 51  \\ 
PQ Gem    & NL/IP  & SP46993  & LP24965 & 0.211636  & -0.5   & 0.00 & low  \\ 
HR Del    & CN     & SP02122  & LR02606 & 0.214165  & -2.6   & 0.17 & 40   \\  
CZ Ori    & DN/UG  & SP07385  &   ---   & 0.214667  & -2.35  & 0.00 & ---    \\ 
TV Col    & NL/IP  & SP08672  & LR07425 & 0.2286    & -2.2   & 0.02 & 70   \\ 
RW Tri    & NL/UX  & SP07915  & LR06895 & 0.231883  & -1.8   & 0.10 & 71  \\ 
VY Scl    & NL/VY  & SP32594  & LP12365 & 0.2323    & -1.9   & 0.06 & 30  \\ 
TX Col    & NL/IP  & SP00502  & LP11930 & 0.23825   & -1.55  & 0.05 & ---  \\ 
RW Sex    & NL/UX  & SP26313  & LP06299 & 0.24507   & -2.0   & 0.02 & 34  \\ 
AH Her    & DN/ZC  & SP17671  & LR13923 & 0.258116  & -2.2   & 0.03 & 46  \\ 
TZ Per    & DN/ZC  & SP17643  & LR13907 & 0.262906  & -2.45  & 0.27 & --- \\ 
TW Pic    & NL/VY  & SP27087  & LR17818 & 0.265     & -2.3   & 0.05 & ---  \\ 
TT Crt    & DN/UG  & SP47801  & LP25680 & 0.2683522 & -2.3   & 0.00 & --- \\ 
SS Cyg    & DN/UG  & SP48441  & LP26199 & 0.27513   & -2.5   & 0.07 & 51  \\  
Z Cam     & DN/ZC  & SP26717  & LP15337 & 0.2898406 & -2.0   & 0.02 & 57  \\ 
EM Cyg    & DN/ZC  & SP07297  & LR06289 & 0.290909  & -2.1   & 0.03 & 67  \\ 
AC Cnc    & NL/UX  & SP18735  & LR14787 & 0.300478  & -1.7   & 0.00 & 76  \\ 
V363 Aur  & NL/UX  & SP25335  & LP05437 & 0.321242  & -1.9   & 0.15 & 70  \\ 
BT Mon    & CN/SW  & SP25446  & LP05516 & 0.333814  & -2.2   & 0.24 & 82   \\  
RZ Gru    & NL/UX  & SP15385  & LR11916 & 0.360     & -1.75  & 0.03 & --- \\ 
RU Peg    & DN/UG  & SP15079  & LR11606 & 0.3746    & -2.1   & 0.00 & 43  \\ 
SY Cnc    & DN/ZC  & SP08082  & LR07051 & 0.38      & -2.3   & 0.00 & --- \\ 
KO Vel    & NL/IP  & SP16034  & LR12341 & 0.422     & -1.4   & 0.05 & ---  \\ 
DX And    & DN/UG  & SP37687  & LP16843 & 0.440502  & -1.9   & 0.10 & 45  \\ 
QU Car    & NL/UX  & SP41926  & LP20699 & 0.45      & -2.3   & 0.11 & $<$60 \\ 
DI Lac    & CN     & SP29325  & LP09208 & 0.543773  & -2.0   & 0.26 & ---  \\   
V841 Oph  & CN     & SP07950  & LR06925 & 0.601304  & -2.6   & 0.40 & --- \\ 
BV Cen    & DN/UG  & SP24867  & LP05162 & 0.610108  & -1.5   & 0.10 & 53  \\ 
GK Per    & NL/IP  & SP20653  & LR16563 & 1.996803  & -0.7   & 0.30 & $<$73  \\ 
\enddata  
\tablecomments{We List here 105 cataclysmic variables for which 
an {\it IUE} spectrum was taken when the system was dominated by 
emission from the disk. 
The CV subtypes are according to \citet{rit03}, except for the
novae which we denote as systems having experienced a 
classical nova explosion (CN). 
The reddening values E(B-V) were taken from \citet{bru94} of directly
assessed based on the 2175\AA\ absorption feature.  
The slope of the UV continuum was assessed for each system
after the spectra were dereddened using the relation from 
\citet{fit07} with R=3.1. 
Periods and inclinations 
are also taken from the Ritter \& Kolb Catalogue \citep{rit03}.}  
\end{deluxetable*}
\end{small}

\begin{figure}[h] 
\vspace{-3.cm} 
\plotone{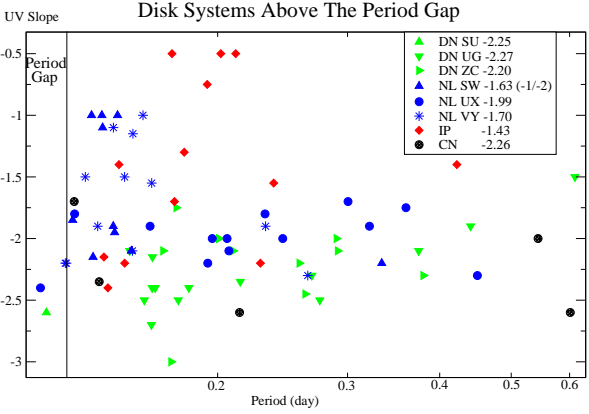}                       
%\vspace{-1.cm} 
\caption{ 
The slope of the ({\it IUE}) UV continuum is shown as a function of the 
orbital period (in days, on a log scale) for the  
disk-dominated CV systems listed in Table 4.
For clarity, only the region above the period gap 
is shown extending to $P \sim 0.6$days. 
The subtypes are shown  as indicated by the
different symbols and colors in the  panel on the upper right, 
where the average value of the slope is also listed for each subtype. 
On the overall DNe SU, UG \& ZC subtypes (all in green) have all a similar 
average slope of about $\sim -2.25$, 
significanlty steeper than the NLs (shown in blue). 
The UX UMa NLs have an average slope
of about -2, the VY Scl NLs have an average slope of -1.70, and the 
eclipsing SW Sex NLs seem to cluster around either -1.0 or -2.0
with an average value of -1.63. 
The intermediate polars  have been included (in red) for comparison 
and have an average slope of only 1.43 (with a very broad scattering)
not inconsistent with magnetically truncated inner disks.   
For the DNe UG \& ZC (as a whole), 
the steepest (most negative) slope value reached at a given 
orbital period shows an increase (less negative) with orbital period,
The SU UMa systems below the gap, shown in detail in 
Fig.20, exhibit a similar pattern. 
We included the novae (CN) classified as IP (CN/IP) with the IPs, 
thereby, leaving only 7 CN systems.  
}
\end{figure}

\begin{figure}[h] 
\vspace{-3.cm} 
\plotone{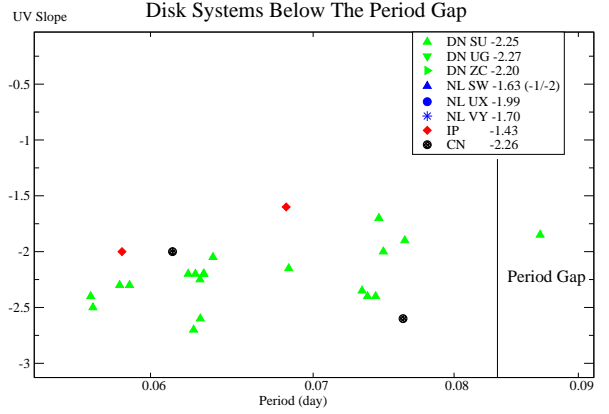}                       
\caption{
The UV slope against the orbital period is shown, as in Fig.19,  
but for the SU UMa systems below the period gap.
The horizontal (log) axis has been streched for clarity.  
As for the the U Gem and Z Cam systems, the  
steepest slope value reached for the SU UMa's at a given 
orbital period shows an increase with orbital period,  
and overal the UV slope value for the SU UMa systems slightly 
increases (less shallow) with orbital period.  
}
\end{figure}

On the overall, the DNe present the steepest average slope (-2.20 to -2.27) 
and, therefore, their disk departs from  the standard disk model  
less than the NLs with an average slope of -1.99 to -1.63.  
Not surprizingly, the intermediate polars, with their magnetically 
truncated inner disks,
have the shallowest average slope of all, with a value of only -1.43,
with a very large scatter ranging from -0.5 to almost -2.5.
This is consistent with the suggestion that the slope of the UV continuum  
becomes shallower as the size of the inner hole in the disk increases. 
We note, however, that the eclipsing SW Sex stars, with an average 
slope value of -1.63, seems to cluster around either -2.0 or -1.0. 

The small number of DNe that are showing soft X-ray emission in outburst  
\citep{van96,kuu06}, 
VW Hyi, OY Car, SW UMa, SS Cyg, U Gem, and Z Cam,
all exhibit a rather steep UV slope: 
-2.4, -2.2, -2.5, -2.5, -2.5, and -2.0 respectively. 
The average value of their UV slope is -2.35, while the
average value of the VY Scl with the hardest X-ray emission 
of all CVs is -1.70, again pointing to the fact that the   
slope of the UV spectral continuum increases with the softness
of the X-ray emission.     

We now consider the UV slope by CV subtype,
excluding the magnetic IPs and eclipsing SW Sex stars, 
we obtain the following  sequence in order of decreasing steepness
of the average UV slope:  
SU UMa and U Gem systems ($\sim -2.26$), 
Z Cam systems ($-2.20$),   
UX UMa systems (-1.99), and 
VY Scl systems (-1.70). 
This sequence is the same for the ratio of the X-ray flux to optical 
flux and/or UV flux of CVs in decreasing order:
from $\sim 0.1$ for SU UMa systems, to U Gem systems, 
to $\sim$0.01 for Z Cam systems, 
to $\sim$0.001 for UX UMa systems and VY Scl systems in high state
\citep{bal15,kuu06}, which is due mainly to an increase of the optical/UV flux
(or $\dot{M}$).  
Within the context of our modified disk model, a simple interpretation
of this correlation is that the  optically thin advective inner disk/BL
region increases with decreasing mass accretion rate.

From Figs. 19 \& 20, we also find that 
for the all the DNe, separately above and below the gap, 
the slope value is {\it possibly increasing slightly}   
(less negative) with increasing orbital period:
the UV slope increases from -2.5 at an orbital period of 
$P \sim 0.15$d to -2.0 at  0.3d$ < P <$0.6d, above the gap (Fig.19);  
and a similar increase is noticable below the period gap,
from -2.5 at $P\sim 0.06$d to -2.0 at $P \sim 0.075$d 
(Fig.20).  
The increase of the UV slope value as a function of the orbital 
period, for the DN systems, is a {\it weak} correlation, 
and it seems to be counter intuitive as one
would expect larger mass accretion rate systems, at longer 
orbital period, to have a disk that is hotter with a steeper
UV slope. 
However, the size of the disk also increases with 
increasing binary separation at longer orbital period, thereby
increasing the surface area of the colder outer disk, providing
a possible explanation for the increase of the UV slope value.     
This is similar to our modified disk modeling, 
where the increasing size and the cold temperature ($\sim 10-12,000$K) 
of the outer disk also makes the UV continuum slope shallower.   

It has been cautioned that systems with a
higher inclination may exhibit a flatter UV continuum
slope \citep{lad91}, however, 
we find that this does not seem to be the case. 
In fact there seems to be a selection effect showing the opposite, 
as most of the VY Scl systems (characterized by a shallow UV continuum
slope)  
in our sample have a low inclination ($< 40$deg) while
most of the remaining non-magnetic systems 
(with a steeper UV continuum slope) in our sample 
have a higher inclination,
making appears as if the low inclination systems have a 
shallow UV slope. 
Many systems in Table 4 do not have a known inclination,
nevertherless, we draw the UV continuum slope against the
inclination in Fig.21. There do not appear to be an obvious 
correlation between the UV slope and the inclination, except 
maybe at a very high inclination, where 3 SW Sex NL system
cluster around -1 (e.g. LX Ser). 
The IPs systems exhibiting the shallowest slope of all (-0.5 to -0.7)  
do not especially have a high inclination: 
FO Aqr has  $i=65$deg and GK Per,$i\sim 46-73$deg,   
but V405 Aur has $i<5$deg, and PQ Gem has a low (but unknown) inclination.
The same is true for the VY Scl systems: 
V425 Cas with a slope of -1.5 has an inclination of only 25deg, 
BZ Cam with a slope of -1.15  has an inclination less than 40deg.

\begin{figure}[h] 
\vspace{-2.cm} 
\plotone{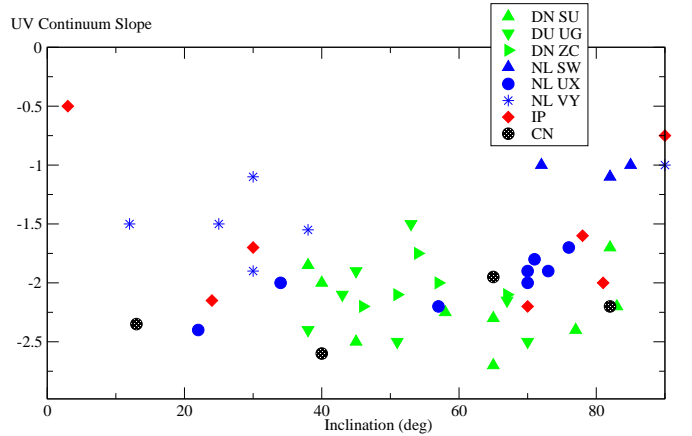}                       
\caption{
The UV slope against the inclination of the system is shown for the
CVs in Table 4 for which the inclination is known. 
Except, possibly, for the few systems close to 90deg, there does 
not appear to be a correlation between the UV slope and the inclination.  
}
\end{figure}

\section{{\bf Summary and Conclusion}}

Our spectral analysis demonstrate that the two VY Scl systems MV Lyr and
BZ Cam, departing from the standard disk model in the UV range 
(with a shallow UV continuum slope of -1.4 and -1.15 respectively),  
can be modeled with a modified/truncated inner disk (of size $\sim R_{\rm wd}$)
and a hot
WD with an inflated radius. 
The UX UMa NL system V592 Cas, with a steeper UV continuum slope 
of -2.4, is modeled with a less modified disk
model with a smaller truncated inner disk region (at $\sim 1.2 R_{\rm wd}$).
The picture we present, to justify our modified (truncated) inner disk, 
is that of a geometrically extended, of size up to $\sim R_{\rm wd}$,  
optically thin BL/inner disk heating the WD through advection of energy,
and which provides an explanation for the observed UV and X-ray 
data in a self-consistent manner.  
This scenario is further strengthened 
by the correlation we found: the non-magnetic disk systems (i.e. excluding
quiescent DNe)  
exhibiting the hardest X-ray (the VY Scl systems in our sample 
in Table 3) have also the the shallowest UV slope, while at the other
end, the disk systems exhibiting soft X-ray (5 DNe in outburst) 
have a much steeper UV slope.

We find that the UV continuum slope of the 
SW Sex systems is possibly affected by their high inclination. 
As to the magnetically truncated inner disk systems, 
the IPs, they have the shallowest 
UV slope of all CVs, consistent with the proposed  modified
disk models with a truncated inner region providing a shallower 
UV continuum slope.

In addition, many NL systems exhibit hot, fast wind outflows 
seen in P Cygni line structure and blue-shifted 
absorption. These winds originate in the inner disk/BL region,
with some systems revealing ejected nebular 
material (e.g.  BZ Cam, \citet{ell84}). 
Also, our X-ray analysis of the NLs MV Lyr, BZ Cam, \& V592 Cas
\citep{bal14},  
suggests that these systems have optically thin BLs possibly merged with 
ADAF-like flows and/or hot coronae. 

The X-ray (hard spectra) and UV characteristics 
(shallow slope) of NL VY systems, including their 
unexpected hot WD \citep{pal17},
can be explained with the presence of 
a geometrically extended optically thin boundary layer/inner disk.
When an accretion flow cannot cool efficiently, due to being optically thin, 
the energy is advected with the flow into the outer layer of the 
star \citep{abr95}, increasing the WD temperature,  
and likely driving an ouflow as in  ADAFs \citep{nar95}. 
For this reason, we expect the VY Scl systems to be modeled with 
the modified disk
model and the addition of a hot WD, possibly with an inflated  
radius. The modeling of such systems has to be carried out by creating
modified disk models suited to each system individually, taking into account 
the WD mass, inner and outer disk radius and the binary separation,
itself depending on the secondary mass and orbital period.   
We note that \citet{bal12} have shown the existence of a truncated inner
disk in quiescent dwarf novae as well, 
where the flow changes from optically thick
into a hot ADAF-like flow. Since quiescent dwarf novae have a very
low $\dot{M}$, the optical thickness in the BL must be very small and 
the truncation radius relatively large. 

The UX UMa systems have an average slope somewhat intermediate between 
that of VY Scl systems and DNe in outburst. Namely, their disk (e.g. V592 Cas) 
is less modified than  for the VY Scl subtype.  
As to the DN systems, as a whole, below and above the period gap, 
they have an average UV continuum slope closest to the standard disk 
and should be modeled with only minimal disk modifications. Their
disk has a geometricallly small and optically thin inner disk/BL region. 
  
We have proposed here a modified disk model compatible with the 
X-ray and UV characteristics of non-magnetic CVs, 
that provides a possible explanation for the  
discrepancy between the data and theory, simultaneously in the 
UV (shallow UV slope) and in the X-ray (optically thin and missing BL). 
We found, by carrying out a quantitative assessment,   
that the UV and X-ray data self-consistently support this scenario, 
and suggest that the
size of the optically thin inner hole (and advection process) in the
disk {\it might be} increasing in the following sequence:
SU UMa and U Gem systems, followed by Z Cam systems, UX UM systems,
and VY Scl systems. In addition, this scenario explains several 
of the characteristics unique to the VY Scl systems, such as their elevated
WD temperature and ejected nebular material.  

To complete this conclusion, we observe that  
the optical spectra of disk-dominated CVs also show a continuum slope
that is rather shallow when compared to theoretical disk spectra,
and the Balmer edge 
(around $\sim$3700-4000\AA ) is often not detected, 
or it is seen in absorption but its size is smaller than expected 
\citep{lad89}.  
It has been suggested that some
of the accretion energy possibly goes in the formation of a disk 
wind contributing to the UV and optical continuum.  
\citet{mat15} have shown that the inclusion of such a disk wind 
significantly improves the spectral fit of the model to the
observed UV and optical spectra and fills up the Balmer edge,  
thereby explaining its non-detection in many optical spectra of 
disk-dominated CVs and providing a viable explanation 
for the observed features of UV and optical spectra in CVs. 
Our present research does not take into account 
emission from a disk wind as it is limited to the emission 
from the optically thick disk.  
In this context our disk model with a truncated inner region  
comes as the disk contribution 
to which a disk wind contribution 
has to be added to provide a more complete picture.
In our modeling (see section 3.7) we include an isothermal outer disk region
with an elevated temperature as the  impinging  of the L1-stream
on the rim of the disk and the tidal interaction from the secondary 
are expected to heat the outer region of the disk \citep{bua01}.
Irradiation from the hot inner boundary layer could also contribute
to heating of the outer disk.  
It is possible that the isothermal outer disk partially  
compensates for the lack of disk wind emission in our modeling. 

%\section*{Acknowledgements} 

We wish to thank The Astrophysical Journal Editor Allen Shafter 
for his promptness in handling the manuscript,  
and in particularly the referee, Ivan Hubeny, 
for his short constructive and encouraging comments and promptness
in delivering a report that helped expedite the article.  
This work was supported by the National Aeronautics and Space 
Administration (NASA) under grant numbers NNX17AF36G, NNX13AF11G \& NNX13AF12G 
issued through the Office of Astrophysics Data Analysis Program (ADAP) 
and grant number NNX13AJ70G issued throught the Astrophysics Division Office
({\it Swift} Cycle 8 Guest Investigator Program) all  
to Villanova University. We have used some of the online data
from the AAVSO, and are thankful to the AAVSO and its members worlwide 
for making this data public and their constant monitoring of cataclysmic 
variables.

\end{document}